\documentclass[aps,twocolumn,superscriptaddress,printnumbers,amsmath,amssymb,10pt]{revtex4-1}
\usepackage[utf8x]{inputenc}
\usepackage{amsmath}
\usepackage{graphicx}
\usepackage[footnotesize,bf]{caption}
\usepackage{subfig}
\usepackage{sidecap}
\usepackage{verbatim} 
\usepackage{microtype}
\usepackage{hyperref}

\def \jcp    {{\it J. Chem. Phys.}}

\begin{document}

\title{Saddle-splay screening and chiral symmetry breaking in toroidal
    nematics}

\author{Vinzenz Koning}
\affiliation{Instituut-Lorentz, Universiteit Leiden, Postbus 9506,
  2300 RA Leiden, The Netherlands}
\email{koning@lorentz.leidenuniv.nl}
\author{Benjamin C. van Zuiden}
\affiliation{Instituut-Lorentz, Universiteit Leiden, Postbus 9506,
  2300 RA Leiden, The Netherlands}
\author{Randall D. Kamien}
\affiliation{Department of Physics and Astronomy, University of
  Pennsylvania, Philadelphia, Pennsylvania 19104, USA}
\author{Vincenzo Vitelli}
\affiliation{Instituut-Lorentz, Universiteit Leiden, Postbus 9506,
  2300 RA Leiden, The Netherlands}

\date{\today}

\begin{abstract}
\noindent We present a theoretical study of director
  fields in toroidal geometries with degenerate planar boundary conditions. We find spontaneous chirality:
  despite the achiral nature of nematics the director configuration show
  a handedness if the toroid is thick enough. In the chiral state the
  director field displays a double twist, whereas in the achiral state
  there is only bend deformation. The critical thickness
  increases as the difference between the twist and saddle-splay moduli grows. A positive
  saddle-splay modulus prefers alignment along the short
  circle of the bounding torus, and hence stimulates promotes a chiral configuration. The chiral-achiral
  transition mimics the order-disorder
  transition of the mean-field Ising
  model. The role of the magnetisation in the Ising model is played by the
  degree of twist. The role of the temperature is played by the
  aspect ratio of the torus.  Remarkably, an external field does not break the
  chiral symmetry explicitly, but shifts the transition. In the case of
  toroidal cholesterics, we do find a preference for one chirality over the other -- the molecular chirality acts as a field in the Ising analogy. 
\end{abstract}

\maketitle

\section{Introduction}
\begin{figure}[h]
    \includegraphics[width=\columnwidth]{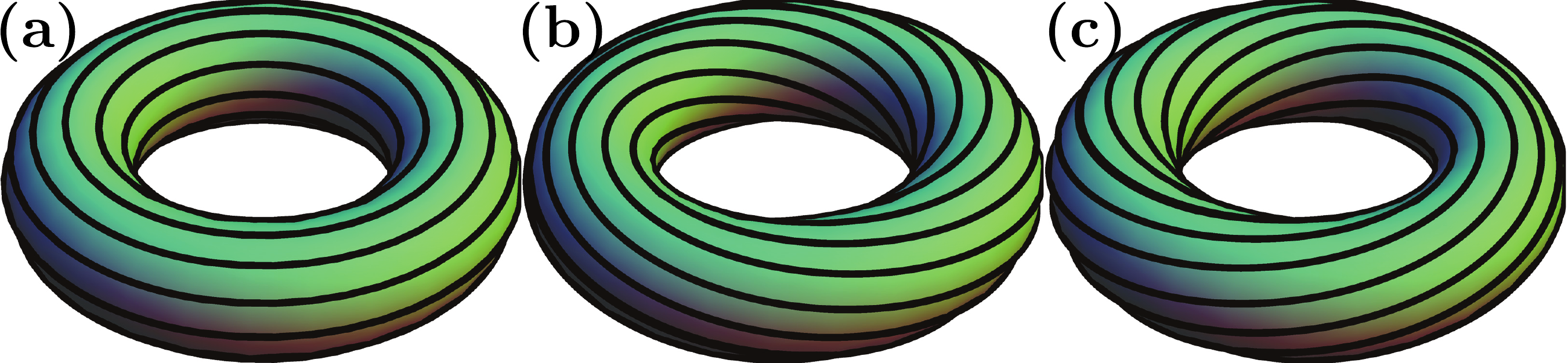}
\caption{\label{fig:chiral} Schematic of (a) achiral, (b) righthanded and (c) lefthanded
toroidal nematic liquid crystals. The black lines are director field lines on the
bounding torus. }
\end{figure}

The confinement of liquid crystals in non-trivial geometries forms a rich and interesting area of
study because
the preferred alignment at the \textit{curved}
bounding surface induce bulk distortions of the liquid crystal -- that is, the boundary conditions {\sl matter}.  This
results in a great diversity of assemblies and mechanical phenomena \cite{Drzaic,doi:10.1080/026782998207640,Poulin199966,Stark2001387,Lopez-Leon:2011fk}. Water droplets dispersed in
a nematic liquid crystal interact and assemble into chains due to the
presence of the anisotropic host fluid \cite{Poulin21031997,PhysRevE.57.610,PhysRevE.57.626}, defect lines in cholesteric
liquid crystals can be knotted and linked around colloidal particles \cite{PhysRevLett.97.127801,PhysRevLett.99.247801,Tkalec01072011,PhysRevE.84.031703}, and surface defects in spherical nematic shells can abruptly
migrate when the thickness inhomogeneity of the shell is
altered \cite{2011NatPh...7..391L,C3SM27671F}. In the examples above
spherical droplets (or colloids), either filled with -- or dispersed in -- a liquid crystal,
create architectures arising from their coupling to the orientational order of
the liquid crystal. Nematic structures where the bounding surface of the colloid or the
liquid crystal droplet is
topologically different from a sphere have also been studied
\cite{Senyuk:2013fk,Liu04062013,C3SM51167G}. Though there has been much interest in the
interplay between order and toroidal
geometries \cite{cond-mat/0012394,PhysRevE.69.041102,0295-5075-67-3-418,PhysRevE.78.010601,2008arXiv0807.4538G,2009AdPhy..58..449B,PhysRevE.87.012603,C3SM51167G},
it was only recently that experimental realisations of nematic liquid crystal
droplets with
toroidal boundaries were reported \cite{Pairam04062013,Smalyukh:2010fk}. Polarised microscopy revealed a twisted nematic orientation in droplets with planar degenerate (tangential) boundary
conditions, despite
the achiral nature of nematics. This phenomenon,
which we will identify as spontaneous chiral symmetry breaking \footnote{Technically, it is spontaneous {\sl achiral} symmetry breaking since the symmetry is the {\sl lack} of chirality.  However, we will conform to the standard convention.}, is
subject of theoretical study in this article. 
The chirality of nematic toroids is
displayed by the the local average orientation of the nematic
molecules, called the director field and indicated by the unit vector $\bf{n}$. Motivated by
experiment, we will assume this
director field to be aligned in the tangent plane of the bounding torus. Fig. \ref{fig:chiral}a
shows an achiral nematic toroid which has its fieldlines
aligned along the azimuthal direction, $\hat{\phi}$. In contrast,
the chiral nematic toroids in Figs. \ref{fig:chiral}b and \ref{fig:chiral}c show a
right and left handedness, respectively, when following the fieldlines anticlockwise
(in the azimuthal direction). 

These nematic toroids share similarites with DNA toroids
\cite{BIP:BIP6}. In fact, twisted DNA toroids have been analysed with
liquid crystal theory \cite{0295-5075-67-3-418,2012EL....10066005S}. Under the appropriate
solvent conditions DNA condenses into toroids \cite{RevModPhys.50.683}. These efficient
packings of genetic material are interesting from a medical viewpoint
as vehicles in therapeutic gene delivery; it has been
argued \cite{0295-5075-67-3-418} that a twist in DNA toroids, for which
there are indicactions both in simulations
\cite{2001BpJ....80..130S,2003JChPh.118.3392S} and experiments \cite{2003PNAS..100.9296C}, would unfold
more slowly and could therefore be
beneficial for this delivery process. 
Thus, besides a way to
engineer complex structures, the theory of geometrically confined
liquid crystals may also provide understanding of biological
systems. 

The organisation of this article is as follows. In section
\ref{sec:director} we will discuss our calculational method which
involves a single variational Ansatz only for the director fields of
both chiral and achiral toroidal nematics. In section
\ref{sec:symmetry_breaking} we will consider its energetics in
relation to the slenderness, elastic anisotropies, cholesteric pitch
and external fields, and discuss the achiral-chiral transition in the
light of the mean field treatment of the Ising model. 
Finally, we
conclude in section \ref{sec:conclusions}.

\section{Toroidal director fields}
\label{sec:director}

\subsection{Free energy of a nematic toroid}
We will study the general case in which the director lies in the tangent plane of the boundary assuming that the anchoring is strong so that the only energy arises from elastic
deformations captured by the
Frank free energy functional \cite{deGennes,SoftMatter}:
\begin{equation}
\label{eq:Frank}
\begin{split}
F[\mathbf{n}\left(\mathbf{x}\right)] &= \frac{1}{2} \int
\text{d}V \left( K_1 \left( \nabla \cdot \bf{n} \right)^2
\right. \\
&+ \left. K_2 \left( \mathbf{n} \cdot \nabla \times \mathbf{n} \right)^2 \right. 
+ \left. K_3 \left( \mathbf{n} \times \nabla \times \mathbf{n}
  \right)^2 \right) \\
&- K_{24} \int \mathbf{dS} \cdot \left( \mathbf{n} \nabla \cdot
  \mathbf{n} +   \mathbf{n} \times \nabla \times \mathbf{n} \right),
  \end{split}
\end{equation}
where $\mathbf{\text{d}S} = \boldsymbol{\nu} \: \text{d}S $ is the area element, with
$\boldsymbol{\nu}$ the unit normal vector and where $\text{d}V$ is the volume element.
Due to the anisotropic nature of the nematic liquid crystal, this
expression contains three bulk elastic moduli, $K_1$, $K_2$, $K_3$, rather than a single one for fully rotationally symmetric
systems. In addition, there is a surface elastic constant
$K_{24}$.  $K_1$, $K_2$, $K_3$ and
$K_{24}$ measure the magnitude of splay, twist, bend and saddle-splay
distortions, respectively.   
We now provide a geometrical
interpretation of the saddle-splay distortions. Firstly, observe that under perfect planar anchoring conditions $\mathbf{n} \cdot
\boldsymbol{\nu}=0$ and so the
first term in the saddle-splay energy does not contribute:\begin{equation}
F_{24} = -K_{24}  \int \text{d}S \; \boldsymbol{\nu} \cdot \left(  \mathbf{n} \times \nabla \times \mathbf{n} \right).
\end{equation}
This remaining term in the saddle-splay energy is often rewritten as
\begin{equation}
F_{24} = K_{24}  \int \text{d}S \; \boldsymbol{\nu} \cdot \left(  \mathbf{n}
  \cdot \nabla  \right) \mathbf{n}.
\end{equation}
because
\begin{align}
 \left( \mathbf{n} \times \nabla \times \mathbf{n} \right)_a
     &=\epsilon_{abc} n_b \epsilon_{cpq} \partial_p n_q \\ \nonumber
& = \left( \delta_{ap} \delta_{bq} - \delta_{aq} \delta_{bp} \right)
n_b \partial_p n_q
\\ & = - n_b \partial_b n_a
\end{align}
where in the last line one uses that  $0 = \partial_a \left( 1\right)
= \partial_a \left( n_b n_b\right)= 2 n_b \partial_a n_b$.  In other words, the bend is precisely the curvature of the integral curves of $\mathbf{n}$.  
Employing the product rule of differention $0=\partial_a \left(\nu_b
  n_b\right) = \nu_b \partial_a n_b + n_b \partial_a \nu_b$
yields 
\begin{equation}
F_{24} = -K_{24}  \int \text{d}S \; \mathbf{n} \cdot\left(  \mathbf{n}
  \cdot \nabla  \right) \boldsymbol{\nu}.
\end{equation}

Upon writing $\mathbf{n} = n_1 \boldsymbol{e}_1 + n_2
\boldsymbol{e}_2$, with $\boldsymbol{e}_1$ and $\boldsymbol{e}_1$ two
orthonormal basis vectors in the plane of the surface, one obtains
\begin{equation}
F_{24} = K_{24}  \int \text{d}S \; n_i L_{ij} n_j,
\end{equation}
where we note that $i,j=1,2$ (rather than running till $3$). Thus the
nematic director couples to the extrinsic curvature tensor \cite{RevModPhys.74.953}, defined as
\begin{equation}
L_{ij} = - \boldsymbol{e}_i \cdot \left( \boldsymbol{e}_j \cdot \nabla
\right) \boldsymbol{\nu}.
\end{equation}
If $\boldsymbol{e}_1$ and $\boldsymbol{e}_2$ are in the directions of
principal curvatures, $\kappa_1$ and $\kappa_2$, respectively, one finds 
\begin{equation}
F_{24} = K_{24}  \int dS \left( \kappa_1  n_1^2 + \kappa_2  n_2^2 \right).
\end{equation}
We conclude that the saddle-splay term favours alignment of the
director along the direction with the smallest principal
curvature if $K_{24}>0$. The controversial surface energy density  $K_{13} \mathbf{n} \nabla \cdot
\mathbf{n}$ is sometimes incorporated in eq. \eqref{eq:Frank}, but is
in our case irrelevant, because the normal vector is perpendicular
to $\mathbf{n}$, and so $\mathbf{n} \cdot \boldsymbol{\nu} =0$. 
 
\begin{figure}[h]
    \includegraphics[width=\columnwidth]{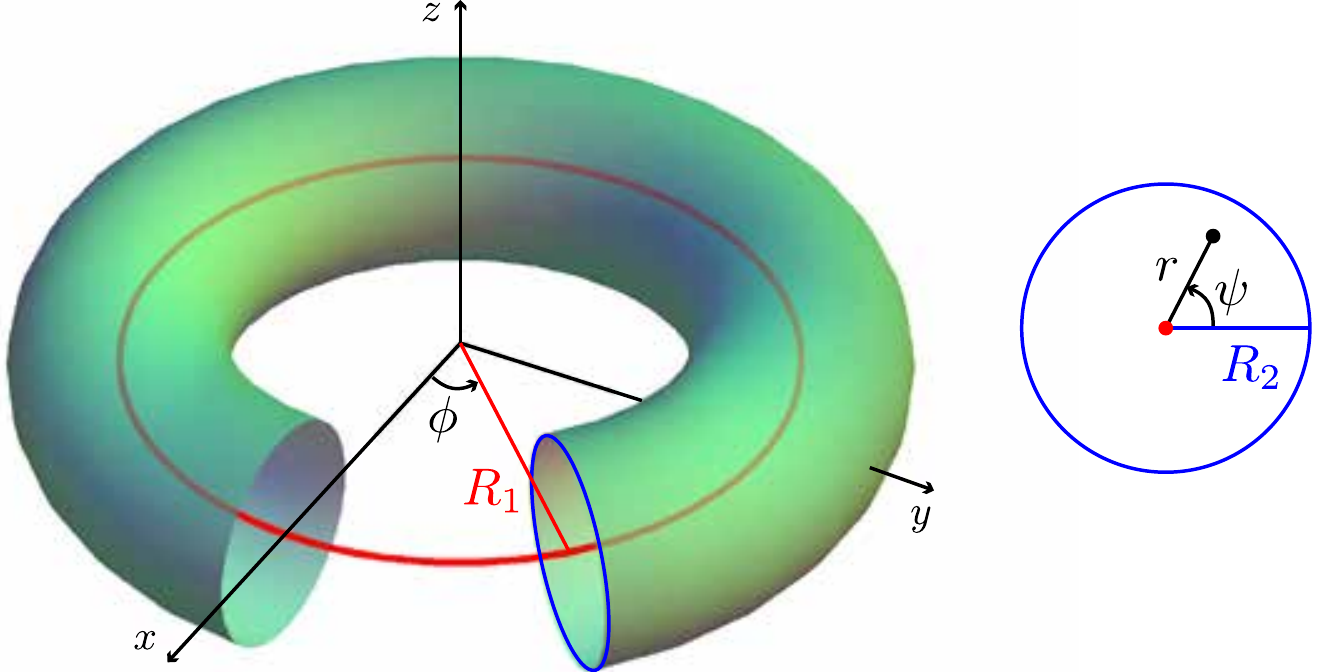}
\caption{\textit{Left panel}: Schematic of the boundary of the
  geometry specified eq. \eqref{eq:implicit} including graphical
  definitions of $\phi$ and $R_1$. The torus characterised by
  a large (red) and a small (blue) circle. The large circle, or centerline, has radius
  $R_1$. \textit{Right
    panel}: Schematic of a cut including graphical definitions of $r$,
  $\psi$ and $R_2$.}
\label{fig:coordinates}
\end{figure}

We will consider a nematic liquid crystal confined in a handle body  bounded by a torus given by the following implicit equation for the cartesian
coordinates $x$, $y$, and $z$:
\begin{equation}
\label{eq:implicit}
\left(R_1 - \sqrt{x^2+y^2} \right)^2 +z^2 \leq R_2^2.
\end{equation}
Here, $R_1$ and $R_2$ are the large and small radii, respectively, of the circles that characterise
the outer surface: a torus obtained by revolving a circle of radius
$R_2$ around the $z$-axis (Fig. \ref{fig:coordinates}). 
We can conveniently parametrise this
solid torus by the coordinates $r\in \left[ 0, R_2 \right]$, $\phi \in
\left[0, 2\pi \right)$ and $\psi \in \left[0, 2\pi \right)$ (illustrated in Fig. \ref{fig:coordinates}):
\begin{align}
x &= \left( R_1 + r \cos \psi  \right)  \cos \phi, \\
y &= \left( R_1 + r \cos  \psi \right) \sin \phi, \\
z &= r \sin \psi.
\end{align}
The metric reads:
\begin{equation}
g_{\mu \nu} =\begin{pmatrix}
1 & 0 &0\\
0 & \left( R_1 + r \cos \psi \right)^2 &0 \\
0 & 0& r^2
 \end{pmatrix},
\end{equation}
with $\mu, \nu \in \{r, \phi, \psi\}$. It follows that $\mathbf{\text{d}S} = \boldsymbol{\nu} \: \sqrt{g} \: \text{d}\psi \, \text{d}\phi$ 
and $\text{d}V = \sqrt{g} \: \text{d}r \, \text{d}\psi \, \text{d}\phi$, where $g=\det g_{\mu \nu}$.

For a torus the $\phi$ and $\psi$ directions are the
principal directions. The curvature along the $\psi$ direction is
everywhere negative and the smallest of the two, so when $K_{24}>0$, the director tends to wind along the small circle with radius $R_2$. 

\subsection{Double twist}
\begin{figure}[h]
\begin{center}
    \includegraphics[width=0.99\columnwidth]{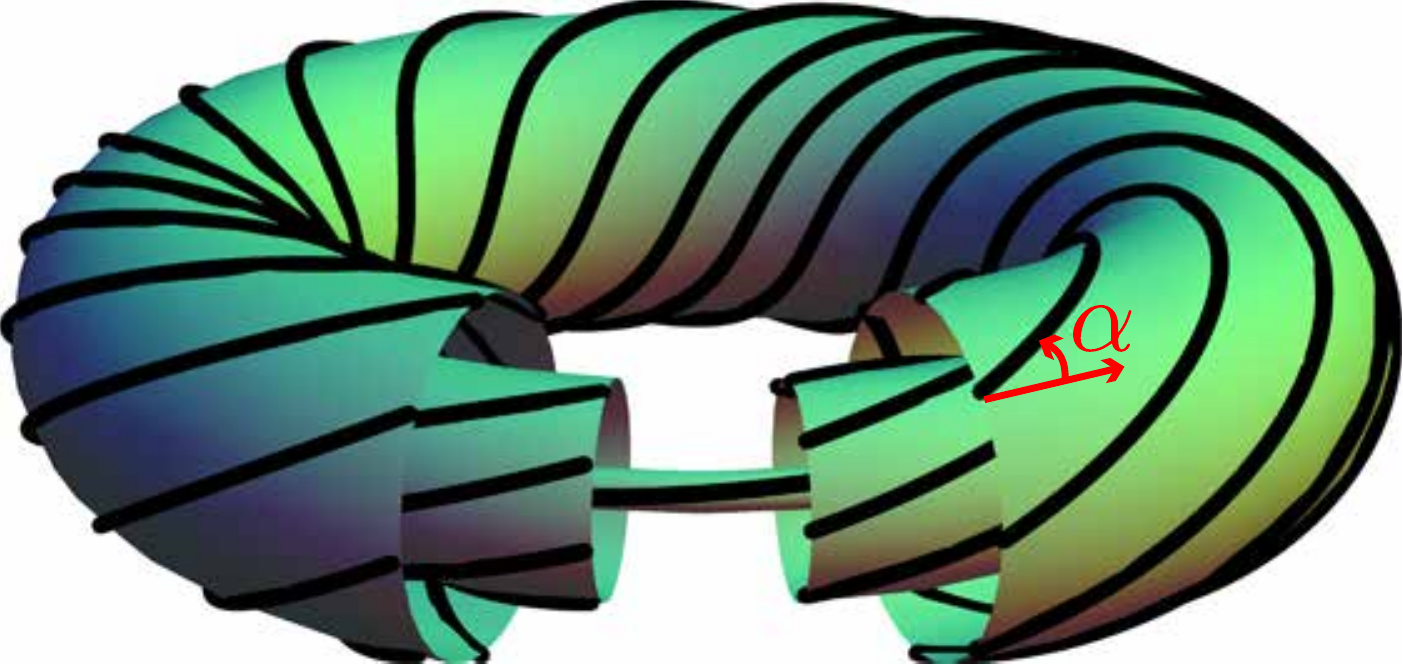}
\end{center}
\caption{Schematic of the Ansatz for the director
fieldlines ($\omega=0.6$ and
$\xi=3$), displaying a twist when going radially outward, including
a graphical definition of $\alpha$.} 
\label{fig:ansatz}
\end{figure}

To minimise the Frank energy we formulate a variational Ansatz built
on several simplifying assumptions \cite{0295-5075-67-3-418}.
We consider a
director field which has no radial component (\textit{i.e.} $n_r=0$), is tangential to the
centerline ($r=0$), and is independent of $\phi$. Furthermore, since we
expect the splay ($K_1$) distortions to be unimportant, we first take the
field to be divergence free ($\textit{i.e.} \nabla \cdot \mathbf{n} = 0$).  Recalling that in curvilinear coordinates the divergence is $\nabla\cdot{\bf n} = \sqrt{g}^{-1} \partial_\mu\left(\sqrt{g}n^\mu\right)$, we write :
\begin{equation}
n_\psi = \frac{ f \left( r \right) R_1 }{\sqrt{g_{\phi \phi}}} 
\end{equation}
where the other terms in $\sqrt{g}$ play no role as they are independent of $\psi$.
The $\phi$-component of the director follows from the normalisation
condition. For the radial
dependence of $f \left(
  r \right)$ we make the simplest choice:
\begin{equation}
f \left( r \right) = \frac{\omega
  r}{R_2}
\end{equation} 
and obtain
\begin{equation}
\label{eq:ansatz}
n_\psi = \omega \frac{ \xi r /R_2 }{\xi + \frac{r}{R_2} \cos \psi},
\end{equation}
where we have introduced $\xi \equiv R_1 / R_2$, the slenderness or
aspect ratio of the torus. 
 The variational parameter $\omega$ governs
the chirality of the toroidal director field.
If $\omega=0$ the director field corresponds to the axial
configuration (Fig \ref{fig:chiral}a). The sign of
$\omega$ determines the chirality: right handed when
$\omega > 0$ (Fig. \ref{fig:chiral}c) and left handed when $\omega < 0$
(Fig. \ref{fig:chiral}b). The magnitude of
$\omega$ determines the degree of twist. Note that the direction of
twist is in the radial direction, as illustrated in Fig. \ref{fig:ansatz}. 
Therefore the toroidal nematic is
doubly twisted, resembling the cylindrical building blocks of the blue
phases \cite{deGennes,SoftMatter}. It may be useful to relate $\omega$ with a quantity at the
surface, say the angle, $\alpha$, that the director makes with
$\hat{\phi}$. For the Ansatz, this angle will be different depending
on whether one measures at the inner or outer part of the torus, but for large $\xi$ we
find
\begin{equation}
\omega \approx n_\psi \biggr\rvert_{r=R_2} \hspace{-12pt} = \sin \alpha. 
\end{equation}

\section{Chiral symmetry breaking}
\label{sec:symmetry_breaking}

\subsection{Results for divergence-free field}
\label{subsec:gammaisone}
\begin{figure}[!t]
\begin{center}
    \includegraphics[width=\columnwidth]{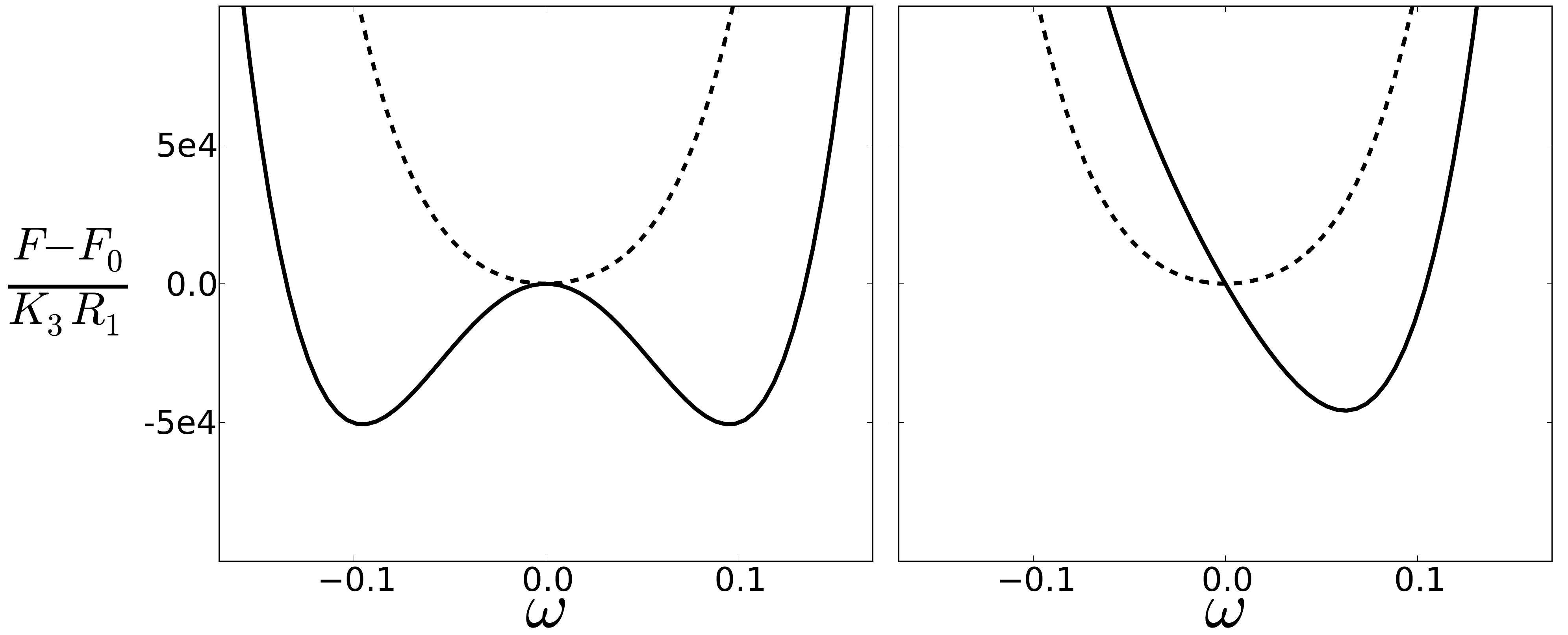}
\end{center}
\caption{\textit{Left panel:} The free energy as a function of
  $\omega$ for $\xi=6$
(dashed) and $\xi=5$ (solid), when $\left(K_2 - K_{24} \right)/K_3 =
10^{-2}$. For $\xi=5$ the chiral symmetry is broken spontaneously: the
minimum values of the energy occurs for a nonzero $\omega$. \textit{Right panel:} The free energy as a function of
  $\omega$ for $q=0$
(dashed) and $q R_2=10^{-3}$ (solid), when $\xi=6$, $\left(K_2 - K_{24} \right)/K_3 =
10^{-2}$ and $K_2/K_3=0.3$. For $q R_2=10^{-3}$ the chiral symmetry is broken explicitly: the
minimum value of the energy occurs for a nonzero $\omega$, because $F$
contains term linear in $\omega$.}
\label{fig:symmetry_breaking}
\end{figure}

\begin{figure}[t]
\begin{center}
    \includegraphics[width=\columnwidth]{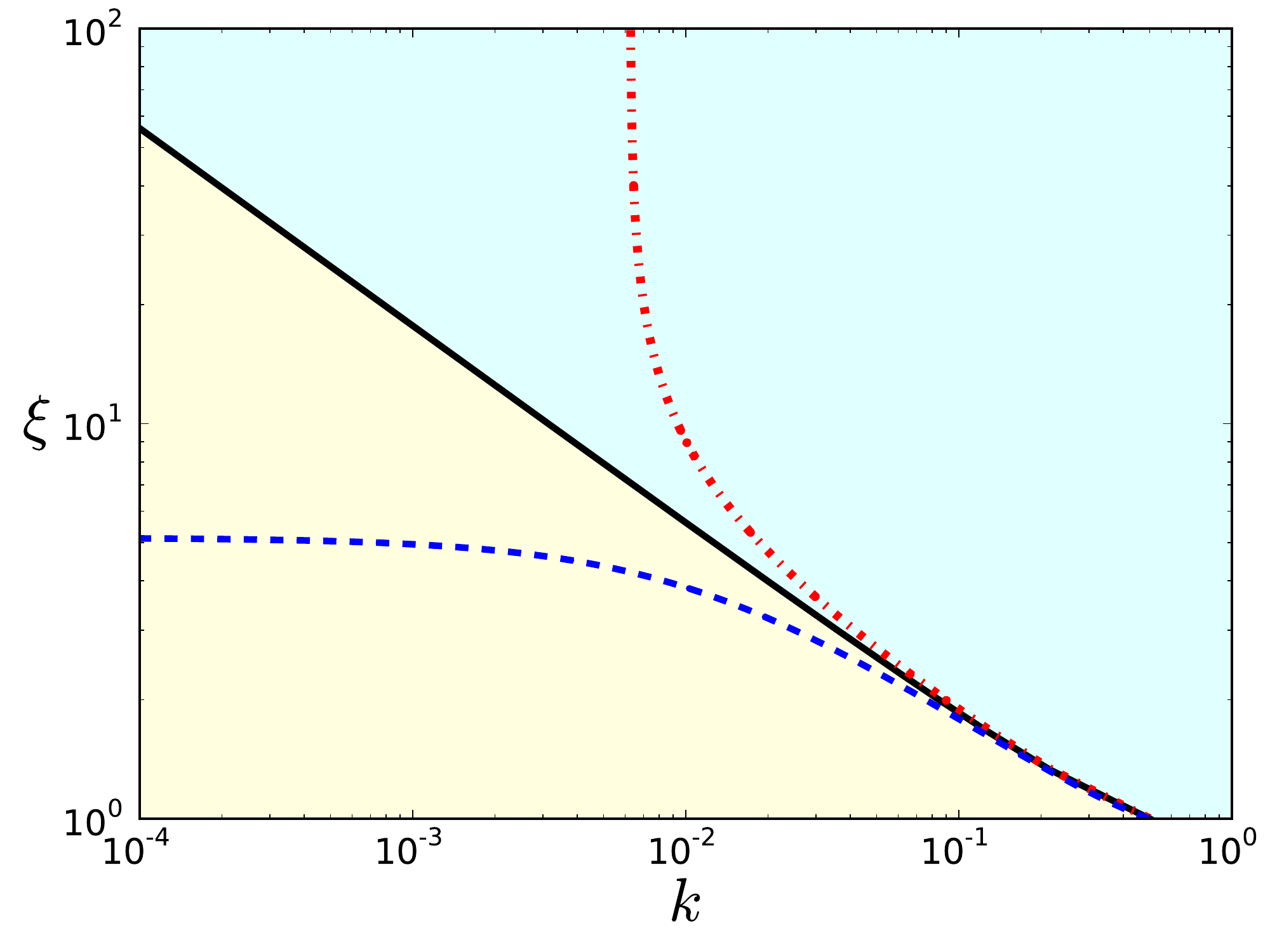}
\end{center}
\caption{Phase diagram as a function of the toroidal slenderness and
the elastic anisotropy in twist and saddle-splay constant,
$k\equiv\left( K_2 - K_{24}\right)/K_3$. The twisted (yellow region)
and axial (cyan region)
configuration are separated by
a boundary line in the absence of an external
field (solid black), when $\mathbf{H}=\sqrt{0.1 K_3} / \left(
  \sqrt{\chi_a} R_2\right)\boldsymbol{\hat{\phi}}$ (dashed blue) and
when $\mathbf{H}=\sqrt{0.1 K_3} / \left(
  \sqrt{\chi_a} R_2\right)\mathbf{\hat{z}}$ (dash-dotted red).}
\label{fig:phase_diagram}
\end{figure}

Since $\omega$ only determines the chirality of the double-twisted
configuration but not the amount of twist,
the free energy is invariant under reversal of the sign of $\omega$,
\textit{i.e.} $F\left(-\omega\right)=F\left(\omega\right)$. This
mirror symmetry allows us to 
write down a Landau-like expansion in which $F$ only contains even
powers of $\omega$,  
\begin{align}
F &= a_0 \left( \{ K_i \}, \xi \right) 
+ a_2 \left( \{
      K_i\}, \xi \right) \omega^2 + a_
4 \left( \{
      K_i\}, \xi \right) \omega^4 \nonumber \\ &+ \mathcal{O} \left( \omega^6 \right)
\end{align}
where $\{K_i\}$ is the set of elastic constants \footnote{Explicitly: $\{K_i\}=\{K_1,K_2,K_3,K_{24}\}$}. If the coefficient $a_2>0$, the achiral nematic toroid ($\omega_{eq}=0$)
corresponds to the minimum of 
$F$ provided that $a_4>0$. In contrast, the mirror symmetry is broken spontaneously
whenever $a_2<0$ (and $a_4>0$). The achiral-chiral critical transition at
$a_2=0$ belongs to the universality class of the mean-field Ising model. Therefore, we can
immediately infer that the value of the critical exponent $\beta$ in
$\omega_{eq} \sim \left(-a_2\right)^\beta$ is $\frac{1}{2}$. To obtain
the dependence of the coefficients $a_i$ on the elastic constants and
$\xi$, we need to evaluate the integral in eq. \eqref{eq:Frank}. We
find for the bend, twist and saddle-splay energies: 
\begin{align}
&\frac{F_3}{K_3 R_1} =   2 \pi^2 \left( \xi -
  \sqrt{\xi^2-1} \right) / \xi  \nonumber \\ \label{eq:bend}&+  \pi^2  \frac{\xi \left( 1 - 9\xi^2 +
    6\xi^4+ 6\xi \sqrt{\xi^2-1} - 6 \xi^3 \sqrt{\xi^2-1}
  \right)}{\left( \xi^2 - 1 \right)^{\frac{3}{2}}}\omega^2
 \nonumber \\ &+\mathcal{O} \left( \omega^4 \right), \\
&\frac{F_{2}}{K_2 R_1} =  4 \pi^2 \frac{\xi^3}{\left(\xi^2
    -1\right)^{\frac{3}{2}}} \omega^2 +\mathcal{O} \left( \omega^6
\right), \\ 
&\frac{F_{24}}{K_{24} R_1} = -4 \pi^2 \frac{\xi^3}{\left(\xi^2
    -1\right)^{\frac{3}{2}}} \omega^2. 
\end{align}
Though the bend and twist energies are Taylor expansions in $\omega$, the
saddle-splay energy is exact. 
The large $\xi$ asymptotic behavior of the elastic energy
reads\footnote{The fourth order term in the bend energy for general $\xi$,
  that reduces to  $\frac{
  \pi^2}{2} K_3 R_2 \xi \omega^4$ in eq. \ref{eq:largexi}, is
  not given in
  eq. \eqref{eq:bend}, because it is too lengthy.}:
\begin{align}
\label{eq:largexi}
\frac{F}{K_3 R_1} \approx \frac{\pi^2}{\xi^2} + 4 \pi^2 \left(k - \frac{5}{16\xi^2} \right) \omega^2 + \frac{\pi^2}{2} \omega^4+ \mathcal{O} \left( \omega^6 \right),
\end{align}
where $k \equiv \frac{K_2-K_{24}}{K_3}$ is the elastic anisotropy
in twist and saddle-splay.
The achiral configuration contains only
bend energy. For sufficiently thick toroids, bend distortions are
exchanged with twist and the mirror symmetry is
indeed broken spontaneously (Fig. \ref{fig:symmetry_breaking}).
Interestingly, if $K_{24}>0$ the saddle-splay deformations screen the cost
of twist. If $K_{24}<0$ on the other hand, there is an extra penalty
for twisting.
Setting the coefficient of the $\omega^2$ term equal to zero yields the phase boundary:
\begin{align}
k_c &= \frac{-1+9\xi_c^2-6\xi_c^4-6\xi_c\sqrt{\xi_c^2 -1} +
  6\xi_c^3\sqrt{\xi_c^2 -1} }{4\xi_c^2} \nonumber \\ &\approx \frac{5}{16 \xi_c^2} \qquad
\text{if} \; \xi \gg 1
\end{align}
Fig.\ref{fig:phase_diagram} shows the phase diagram as a function of $\xi$ and
$k$. 
It is interesting to look at the critical
behavior. The degree of twist close to the transition is
\begin{equation}
\alpha_{eq} \approx \omega_{eq} \approx 2 \left( \frac{5}{16\xi^2} - k\right)^{1/2}
\end{equation}
where we have used that $\sin \alpha_{eq} \approx \alpha_{eq}$ for
small $\alpha_{eq}$. Upon expanding $\xi=\xi_c + \delta \xi$ (with
$\delta \xi <0$) and $k=k_c + \delta k$  (with $\delta k <0$) around their critical
values $\xi_c$ and $k_c$, respectively, we obtain the following
scaling relations:
\begin{align}
\label{eq:deltaxi}
\alpha_{eq} &\approx \frac{\sqrt{5}}{2}\left(-\frac{ \delta
    \xi}{\xi_c^3} \right)^{1/2} \\ 
\label{eq:deltak}
\alpha_{eq} &\approx 2\left(-\delta k\right)^{1/2} 
\end{align}
while keeping $k$ and $\xi$ fixed,
respectively. Eqs. \ref{eq:deltaxi} and \ref{eq:deltak} are analogues
to $m_{eq} \sim \left( -t\right)^{1/2}$, relating the equilibrium
magnetisation, $m_{eq}$ (in the ferromagnetic
phase of the Ising model in Landau theory), to the reduced
temperature, $t$.

\subsection{Effects of external fields and cholesteric pitch}
Due to the inversion symmetry of nematics, $F\left[\mathbf{n}\right] = F\left[-\mathbf{n}\right]$,
an external magnetic field, $\mathbf{H}$, couples quadratically to the
components of $\mathbf{n}$
rather than linearly as in spin systems. The magnetic free energy
contribution reads:
\begin{equation}
F_m = -\frac{\chi_a}{2} \int \text{d}V \left( \mathbf{n} \cdot \mathbf{H}\right)^2,
\end{equation}
where $\chi_a = \chi_\parallel - \chi_\perp$, the difference between the magnetic
susceptibilities parallel and perpendicular to $\mathbf{n}$.
Consequently, there is no explicit chiral
symmetry breaking due to $\mathbf{H}$ as is the case in the Ising model. Rather, $\mathbf{H}$ shifts the
location of the critical transition in the phase diagram. For
concreteness, we will consider two different applied fields, namely
a uniaxial field $\mathbf{H} = H_z \mathbf{\hat{z}}=H_z\sin (\psi)
\mathbf{\hat{r}}+H_z \cos (\psi)\boldsymbol{\hat{\psi}}$ and an azimuthal
field $\mathbf{H} = H_\phi
\boldsymbol{\hat{\phi}}$, as if produced by a conducting wire going
through the hole of the toroid. For $\mathbf{H} = H_z
\mathbf{\hat{z}}$ we find
\begin{align}
\label{eq:F_with_field_z}
F_m &= - \pi^2 \chi_a H_z^2 R_1 R_2^2 \xi^2 \left( 2 \xi \left( \xi
    -\sqrt{\xi^2-1}\right) -1 \right) \omega^2 \nonumber \\ 
&\approx - \frac{\pi^2}{4} \chi_a H_z^2 R_1 R_2^2 \omega^2 \qquad
\text{if} \; \xi \gg 1. 
\end{align}
For a positive $\chi_a$ this energy contribution is negative, implying
that a larger area in the phase diagram is occupied by the twisted
configuration. The new phase boundary (Fig. \ref{fig:phase_diagram}), which is now a surface in the
volume spanned by $\xi$, $k$ and $H_z$ instead of a line, reads:
\begin{align}
k_c &= \left[-1+9\xi_c^2-6\xi_c^4-6\xi_c\sqrt{\xi_c^2 -1} +
  6\xi_c^3\sqrt{\xi_c^2 -1} \right. \nonumber \\ 
& -\frac{\chi_a \left(H_z\right)_c^2 R_2^2}{K_3}
\left(\xi_c^2-1\right) \xi_c \nonumber \\
&\times \left. \left( -2\xi_c+2\xi_c^3+\sqrt{\xi_c^2-1}-2\xi_c^2\sqrt{\xi_c^2-1}\right)\right]
/ \left(4\xi_c^2\right) \nonumber \\ 
& \approx \frac{5}{16 \xi_c^2} + \frac{ \chi_a \left(H_z\right)_c^2 R_2^2}{16K_3}\quad
\text{if} \; \xi \gg 1.
\end{align}
In contrast, an azimuthal field favours the axial configuration,
contributing a postive $\omega^2$-term to the energy when $\chi_a>0$:
\begin{align}
&F_m = -\pi^2 \chi_a H_\phi^2 R_1 R_2^2 \nonumber \\ 
&+ \frac{2\pi^2}{3} \chi_a H_\phi^2 R_1 R_2^2 \xi \left( 2 \xi^2 \left( \xi
    -\sqrt{\xi^2-1}\right) -\sqrt{\xi^2-1} \right) \omega^2 \nonumber \\ 
&\approx -\pi^2 \chi_a H_\phi^2 R_1 R_2^2 + \frac{\pi^2}{2} \chi_a H_\phi^2 R_1 R_2^2 \omega^2 \quad
\text{if} \; \xi \gg 1. 
\end{align}
Consequently, this yields a shifted phase boundary (Fig. \ref{fig:phase_diagram}):
\begin{align}
k_c &= \left[-1+9\xi_c^2-6\xi_c^4-6\xi_c\sqrt{\xi_c^2 -1} +
  6\xi_c^3\sqrt{\xi_c^2 -1} \right. \nonumber \\ 
& -\frac{2\chi_a \left(H_\phi\right)_c^2 R_2^2}{3K_3} \left(\xi_c^2-1\right) \nonumber \\
&\times \left. \left( 1+\xi_c^2-2\xi_c^4+2\xi_c^3\sqrt{\xi_c^2-1}\right)\right]
/ \left(4\xi_c^2\right) \nonumber \\ 
& \approx \frac{5}{16 \xi_c^2} - \frac{ \chi_a \left(H_\phi\right)_c^2 R_2^2}{8K_3} \qquad
\text{if} \; \xi \gg 1.
\label{eq:phase_boundary_with_field_phi}
\end{align}
Similar results (eqs. \eqref{eq:F_with_field_z} to \eqref{eq:phase_boundary_with_field_phi}) hold for an applied electric field $\mathbf{E}$
instead of a magnetic field; the analog of $\chi_a$ is the dielectric anisotropy. There could however be
another physical mechanism at play in a nematic insulator, namely the
flexoelectric effect \cite{PhysRevLett.22.918, deGennes}. Splay and bend deformations induce a
polarisation
\begin{equation}
\label{eq:polarisation}
\mathbf{P} = e_1 \mathbf{n}  \nabla \cdot \mathbf{n}  +
e_3  \mathbf{n} \times \nabla \times \mathbf{n},
\end{equation}
where $e_1$ and $e_3$ are called the flexoelectric coefficients. Note
that the first term in eq. \ref{eq:polarisation} is irrelevant for
the divergence-free Ansatz. 
A coupling of $\mathbf{P}$ with $\mathbf{E}$
\begin{equation}
\label{eq:Fpol}
F_P = -\int dV \mathbf{P} \cdot \mathbf{E}
\end{equation}
could potentially lead to a shift of the transition. In the particuar
case when $\mathbf{E} =
E_z \mathbf{\hat{z}} = E_z\sin (\psi)
\mathbf{\hat{r}}+E_z \cos (\psi)\boldsymbol{\hat{\psi}}$, however, the $\omega^2$ contribution from
eq. \eqref{eq:Fpol} vanishes, thus not yielding such a shift.

If we now consider toroidal cholesterics rather than nematics, the chiral
symmetry is broken explicitly (Fig. \ref{fig:symmetry_breaking}). A cholesteric pitch of $2\pi / q$
gives a contribution to the free energy of:
\begin{equation}
F_{cn} = K_2 \, q \int \text{d}V ~\: \mathbf{n} \cdot \nabla \times \mathbf{n}.
\end{equation}
Substituting eq. \ref{eq:ansatz} yields
\begin{align}
F_{cn} &= -8 \pi^2 K_2 \, q  \, R_1 R_2 \,\xi \left( \xi - \sqrt{\xi^2
    -1}\right) \omega  + \mathcal{O} \left( \omega^3 \right)\nonumber \\
&\approx -4 \pi^2 K_2 \, q \, R_1 R_2\, \omega + \mathcal{O} \left( \omega^3 \right) \qquad
\text{if} \; \xi \gg 1.
\end{align}
Therefore, at the critical line in the phase diagram spanned by $k$
and $\xi$, the degree of twist or surface angle scales (for large $\xi$) with the helicity of the
cholesteric as
\begin{equation}
\alpha_{eq} \approx \left( 2K_2R_2\,q/K_3\right)^{1/3} \sim q^{1/3}.
\end{equation}
This is the analog scaling relation of $m_{eq}\sim H^{1/3}$ in the
mean-field Ising model.

\subsection{Results for the two-parameter Ansatz}
\label{subsec:generalgamma}
Motivated by experiments \cite{Pairam04062013}, we can introduce an extra variational
paramer $\gamma$ to allow for splay deformations, in addition to $\omega$: 
\begin{equation}
\label{eq:ansatz2}
n_\psi = \omega \frac{ \xi r /R_2 }{\xi + \gamma\frac{r}{R_2} \cos \psi}.
\end{equation}
(Note that eqn \ref{eq:ansatz} is recovered by setting $\gamma=1$ in
eqn \ref{eq:ansatz2}.)
In subsection \ref{subsec:gammaisone} analytical results for
$\gamma=1$ were presented. In this subsection we will slightly improve these
results by finding the optimal value of $\gamma$ numerically. Firstly,
we discretise the azimuthally symmetric director field in the $r$ and $\psi$ direction. Next, we compute the Frank free energy density (eq. \ref{eq:Frank}) by taking
finite differences \cite{fornberg1988generation} of the discretised nematic field. After summation over the
volume elements the Frank free energy will become a
function of $\omega$ and $\gamma$ for a given set of elastic constants
and a given aspect ratio. Because of the normalisation condition on
$\mathbf{n}$, the allowed values for $\omega$ and $\gamma$
are constrained to the open diamond-like interval for which $-\xi <
\gamma < \xi$ and $\tfrac{\left|\gamma\right|-\xi}{\xi} < \omega <
\tfrac{\xi-\left|\gamma\right|}{\xi}$ holds. 

\begin{figure}[h]
    \subfloat[]
    {
    	\label{fig:gamma-alpha}
    	\includegraphics[width=\columnwidth]{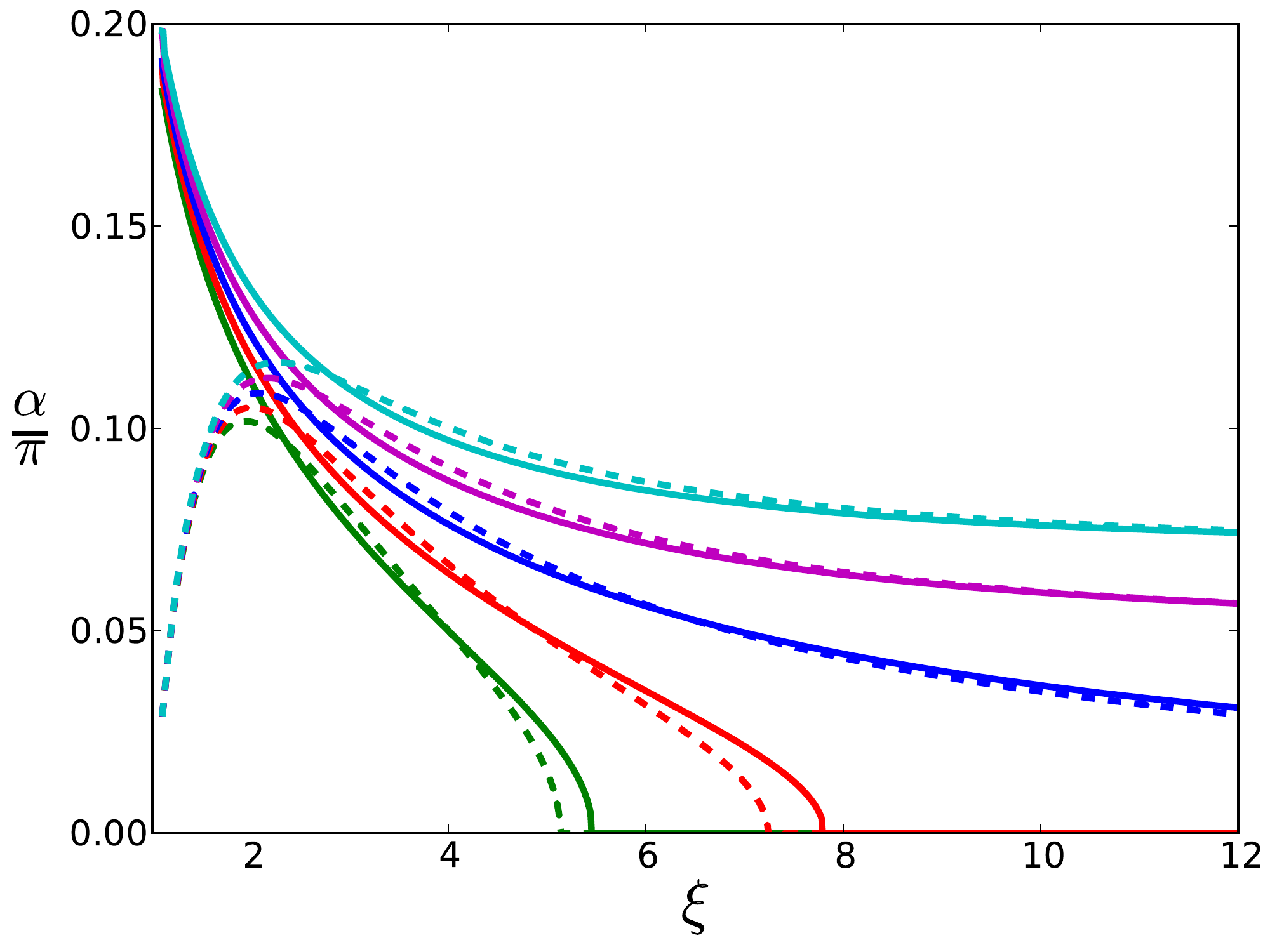}
    } 
    \\
    \subfloat[]
    {
    	\label{fig:phase} 
    	\includegraphics[width=\columnwidth]{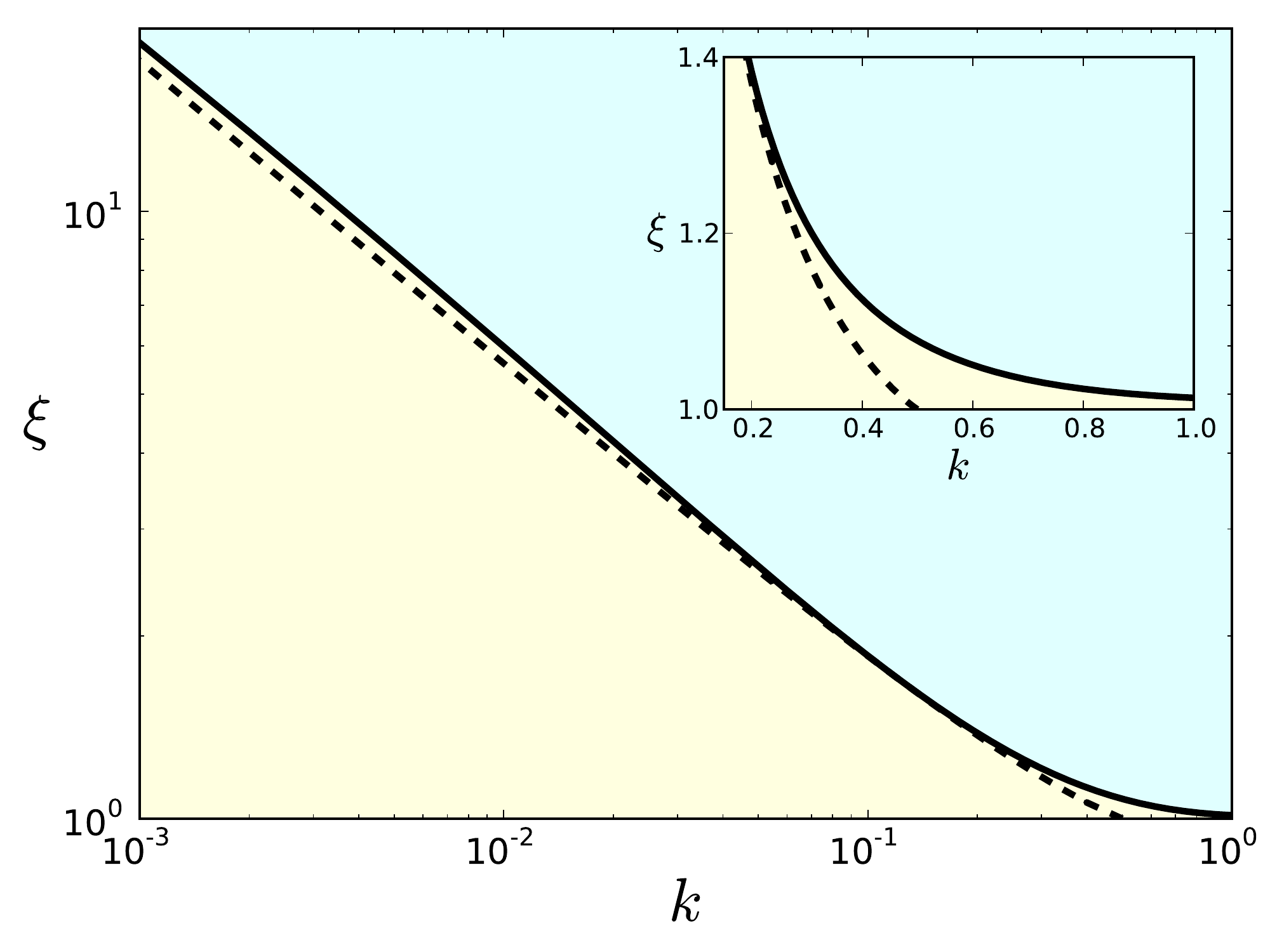}
	}
	\caption{{\bf(a)} Twist angle $\alpha$ (in units of $\pi$) at $\psi = \pi /2$ versus the slenderness $\xi$ for $k=0.012$ (green), $k=0.006$ (red), $k=0$ (blue), $k=-0.006$ (magenta) and $k=-0.012$ (cyan). The dashed lines represent $\alpha$ for $\gamma=1$, the solid lines represent $\alpha$ found for the optimal $\gamma$. {\bf(b)} The phase boundary as a function of the toroidal slenderness $\xi$ and elastic anisotropy $k$ for $\gamma$ as a variational parameter(solid) and for $\gamma=1$ (dashed). The inset zooms in on the phase boundary for small $\xi$.}
\end{figure}

The minima of the energy surface can be found by employing the conjugate
gradient method. We have looked at the difference
between the $\gamma=1$ case and the case where the value of $\gamma$ is
chosen to minimise the energy. This was done for various choices of
$k$. We have chosen the material properties of 5CB, \textit{i.e.} $K_1
= 0.64 K_3$ and $K_2 = 0.3K_3$ \cite{SoftMatter}. The value for of
$K_{24}$ has not been so accurately determined, but previous measurements 
\cite{PhysRevLett.67.1442,PhysRevLett.73.979,PhysRevE.49.R978,PhysRevE.49.1344,doi:10.1142/S0217979295000926,Pairam04062013}
seem to suggest that $K_{24} \approx K_2$, corresponding to $k \approx
0$. 

We are interested in how the phase boundary changes by introducing the variational parameter $\gamma$. Therefore, the twist angle $\alpha$, evaluated at the surface of the torus at $\psi = \frac{\pi}{2}$, versus the slenderness $\xi$ is shown in Fig. \ref{fig:gamma-alpha}. For the particular choices of $k$ there are two noticeable differences between the single-parameter ansatz and the two-parameter ansatz. Firstly, for small values of $\xi$, $\alpha$ is changed significantly. Secondly, for larger values of $\xi$ we see that if there is a chiral-achiral phase transition, it is shifted by a small value with respect to $\xi$. In Fig. \ref{fig:phase} we further investigate how introducing $\gamma$ influences the phase boundary, by plotting the phase boundary as a function of the toroidal slenderness $\xi$ and elastic anisotropy $k$ for both $\gamma$ as a variational parameter (solid) and for $\gamma=1$ (dashed). Observe that, for both the small $\xi$ and small $k$ regime, the difference is significant.

\section{Conclusions}
\label{sec:conclusions}
We have investigated spontaneous chiral symmetry breaking in toroidal
nematic liquid crystals. As in the case of nematic tactoids
\cite{0953-8984-16-49-003,Tortora29032011}, the two ingredients for this macroscopic chirality
are orientational order of achiral microscopic constituents and a
curved confining boundary. This phenomenon occurs when both the aspect
ratio of the toroid and $\frac{K_2-K_{24}}{K_3}$ are small.
The critical behavior of the transition belongs to the same universality class as the
ferromagnet-paramagnet transition in the Ising
model in dimensions above the upper critical dimension. The analogues of the
magnetisation, reduced temperature and external field are
the degree of twist (or surface angle), slenderness or
$\frac{K_2-K_{24}}{K_3}$, and (cholesteric) helicity 
in liquid crystal toroids, respectively. Critical exponents are collected
in Table \ref{table:universality}. 

Thus, the helicity rather than an external field breaks
the chiral symmetry explicitly. Remarkably, since an external field couples quadratically
to the director field, it induces a shift of the phase boundary. An
azimuthally aligned field favours the mirror symmetric director configuration, whereas a
homogeneous field in the $z$-direction favours the doubly twisted configuration. 

Finally, it is interesting to note that experimental measurements of
the twist angle versus the reduced temperature (\textit{i.e.}
difference in temperature and the transition temperature) in spherical bipolar droplets yield
an exponent of $0.75 \pm 0.1$ and $0.76 \pm 0.1$ for 8\,CB and 8\,OCB,
respectively \cite{Lavrentovich:1990fk}. Since the same line of reasoning outlined in
this article for toroidal droplets applies to spherical droplets \cite{0305-4470-19-16-019} and there is no
non-analytic behavior of the elastic constants as a function of
temperature, one would have expected and exponent of $\tfrac{1}{2}$. This
discrepancy is puzzling and we are very keen on seeing experimental studies
of the critical behavior of the chiral-achiral transition in nematic
toroids.

\begin{table}[h]
\caption{Dictionary of liquid crystal toroid and mean-field Ising model.} 
\centering 
\begin{tabular}{c c c } 
\hline \hline \noalign{\smallskip} 
Liquid crystal toroid & Mean-field Ising model & Exponent  \\ [0.5ex] 
\hline  \noalign{\smallskip}
$\alpha_{eq} \sim \left( -\delta \xi \right)^\beta$ & $m_{eq} \sim \left(
  -t\right)^\beta$ & $\beta = 1/2$ \\ [0.5ex]
$\alpha_{eq} \sim \left( -\delta k \right)^\beta$ & \  & \  \\ [0.5ex]
$\alpha_{eq} \sim q^{1/\delta}$ & $m_{eq} \sim H^{1/\delta}$  & $\delta = 3$ \\
[0.5ex] 
\hline \hline 
\end{tabular}
\label{table:universality} 
\end{table}

\section*{Acknowledgement}
V.K. acknowledges funding from Stichting Fundamenteel Onderzoek der
Materie (FOM). This work was partially supported by NSF DMR12-62047
and a Simons Investigator award from the Simons Foundation to RDK. We
would like to thank Hiroshi Yokoyama for discussions.
\bibliography{liquid_crystals} 

\begin{thebibliography}{49}%
\makeatletter
\providecommand \@ifxundefined [1]{%
 \@ifx{#1\undefined}
}%
\providecommand \@ifnum [1]{%
 \ifnum #1\expandafter \@firstoftwo
 \else \expandafter \@secondoftwo
 \fi
}%
\providecommand \@ifx [1]{%
 \ifx #1\expandafter \@firstoftwo
 \else \expandafter \@secondoftwo
 \fi
}%
\providecommand \natexlab [1]{#1}%
\providecommand \enquote  [1]{``#1''}%
\providecommand \bibnamefont  [1]{#1}%
\providecommand \bibfnamefont [1]{#1}%
\providecommand \citenamefont [1]{#1}%
\providecommand \href@noop [0]{\@secondoftwo}%
\providecommand \href [0]{\begingroup \@sanitize@url \@href}%
\providecommand \@href[1]{\@@startlink{#1}\@@href}%
\providecommand \@@href[1]{\endgroup#1\@@endlink}%
\providecommand \@sanitize@url [0]{\catcode `\\12\catcode `\$12\catcode
  `\&12\catcode `\#12\catcode `\^12\catcode `\_12\catcode `\%12\relax}%
\providecommand \@@startlink[1]{}%
\providecommand \@@endlink[0]{}%
\providecommand \url  [0]{\begingroup\@sanitize@url \@url }%
\providecommand \@url [1]{\endgroup\@href {#1}{\urlprefix }}%
\providecommand \urlprefix  [0]{URL }%
\providecommand \Eprint [0]{\href }%
\@ifxundefined \urlstyle {%
  \providecommand \doi  [0]{\begingroup \@sanitize@url \@doi}%
  \providecommand \@doi [1]{\endgroup \@@startlink {\doibase
  #1}doi:\discretionary {}{}{}#1\@@endlink }%
}{%
  \providecommand \doi  [0]{doi:\discretionary{}{}{}\begingroup
  \urlstyle{rm}\Url }%
}%
\providecommand \doibase [0]{http://dx.doi.org/}%
\providecommand \Doi [0]{\begingroup \@sanitize@url \@Doi }%
\providecommand \@Doi  [1]{\endgroup\@@startlink{\doibase#1}\@@Doi}%
\providecommand \@@Doi [1]{#1\@@endlink}%
\providecommand \selectlanguage [0]{\@gobble}%
\providecommand \bibinfo  [0]{\@secondoftwo}%
\providecommand \bibfield  [0]{\@secondoftwo}%
\providecommand \translation [1]{[#1]}%
\providecommand \BibitemOpen [0]{}%
\providecommand \bibitemStop [0]{}%
\providecommand \bibitemNoStop [0]{.\EOS\space}%
\providecommand \EOS [0]{\spacefactor3000\relax}%
\providecommand \BibitemShut  [1]{\csname bibitem#1\endcsname}%
\bibitem [{\citenamefont {Drzaic}(1995)}]{Drzaic}%
  \BibitemOpen
  \bibfield  {author} {\bibinfo {author} {\bibfnamefont {P.~S.}\ \bibnamefont
  {Drzaic}},\ }\href@noop {} {\emph {\bibinfo {title} {Liquid crystal
  dispersions}}}\ (\bibinfo  {publisher} {World Scientific},\ \bibinfo {year}
  {1995})\BibitemShut {NoStop}%
\bibitem [{\citenamefont {Lavrentovich}(1998)}]{doi:10.1080/026782998207640}%
  \BibitemOpen
  \bibfield  {author} {\bibinfo {author} {\bibfnamefont {O.~D.}\ \bibnamefont
  {Lavrentovich}},\ }\Doi {10.1080/026782998207640} {\bibfield  {journal}
  {\bibinfo  {journal} {Liquid Crystals},\ }\textbf {\bibinfo {volume} {24}},\
  \bibinfo {pages} {117} (\bibinfo {year} {1998})}\BibitemShut {NoStop}%
\bibitem [{\citenamefont {Poulin}(1999)}]{Poulin199966}%
  \BibitemOpen
  \bibfield  {author} {\bibinfo {author} {\bibfnamefont {P.}~\bibnamefont
  {Poulin}},\ }\Doi {10.1016/S1359-0294(99)00009-6} {\bibfield  {journal}
  {\bibinfo  {journal} {Current Opinion in Colloid \& Interface Science},\
  }\textbf {\bibinfo {volume} {4}},\ \bibinfo {pages} {66 } (\bibinfo {year}
  {1999})},\ ISSN \bibinfo {issn} {1359-0294}\BibitemShut {NoStop}%
\bibitem [{\citenamefont {Stark}(2001)}]{Stark2001387}%
  \BibitemOpen
  \bibfield  {author} {\bibinfo {author} {\bibfnamefont {H.}~\bibnamefont
  {Stark}},\ }\Doi {10.1016/S0370-1573(00)00144-7} {\bibfield  {journal}
  {\bibinfo  {journal} {Physics Reports},\ }\textbf {\bibinfo {volume} {351}},\
  \bibinfo {pages} {387 } (\bibinfo {year} {2001})},\ ISSN \bibinfo {issn}
  {0370-1573}\BibitemShut {NoStop}%
\bibitem [{\citenamefont {Lopez-Leon}\ and\ \citenamefont
  {Fernandez-Nieves}(2011)}]{Lopez-Leon:2011fk}%
  \BibitemOpen
  \bibfield  {author} {\bibinfo {author} {\bibfnamefont {T.}~\bibnamefont
  {Lopez-Leon}}\ and\ \bibinfo {author} {\bibfnamefont {A.}~\bibnamefont
  {Fernandez-Nieves}},\ }\Doi {10.1007/s00396-010-2367-7} {\bibfield  {journal}
  {\bibinfo  {journal} {Colloid and Polymer Science},\ }\textbf {\bibinfo
  {volume} {289}},\ \bibinfo {pages} {345} (\bibinfo {year} {2011})},\ ISSN
  \bibinfo {issn} {0303-402X}\BibitemShut {NoStop}%
\bibitem [{\citenamefont {Poulin}\ \emph {et~al.}(1997)\citenamefont {Poulin},
  \citenamefont {Stark}, \citenamefont {Lubensky},\ and\ \citenamefont
  {Weitz}}]{Poulin21031997}%
  \BibitemOpen
  \bibfield  {author} {\bibinfo {author} {\bibfnamefont {P.}~\bibnamefont
  {Poulin}}, \bibinfo {author} {\bibfnamefont {H.}~\bibnamefont {Stark}},
  \bibinfo {author} {\bibfnamefont {T.~C.}\ \bibnamefont {Lubensky}}, \ and\
  \bibinfo {author} {\bibfnamefont {D.~A.}\ \bibnamefont {Weitz}},\ }\Doi
  {10.1126/science.275.5307.1770} {\bibfield  {journal} {\bibinfo  {journal}
  {Science},\ }\textbf {\bibinfo {volume} {275}},\ \bibinfo {pages} {1770}
  (\bibinfo {year} {1997})}\BibitemShut {NoStop}%
\bibitem [{\citenamefont {Lubensky}\ \emph {et~al.}(1998)\citenamefont
  {Lubensky}, \citenamefont {Pettey}, \citenamefont {Currier},\ and\
  \citenamefont {Stark}}]{PhysRevE.57.610}%
  \BibitemOpen
  \bibfield  {author} {\bibinfo {author} {\bibfnamefont {T.~C.}\ \bibnamefont
  {Lubensky}}, \bibinfo {author} {\bibfnamefont {D.}~\bibnamefont {Pettey}},
  \bibinfo {author} {\bibfnamefont {N.}~\bibnamefont {Currier}}, \ and\
  \bibinfo {author} {\bibfnamefont {H.}~\bibnamefont {Stark}},\ }\Doi
  {10.1103/PhysRevE.57.610} {\bibfield  {journal} {\bibinfo  {journal} {Phys.
  Rev. E},\ }\textbf {\bibinfo {volume} {57}},\ \bibinfo {pages} {610}
  (\bibinfo {year} {1998})}\BibitemShut {NoStop}%
\bibitem [{\citenamefont {Poulin}\ and\ \citenamefont
  {Weitz}(1998)}]{PhysRevE.57.626}%
  \BibitemOpen
  \bibfield  {author} {\bibinfo {author} {\bibfnamefont {P.}~\bibnamefont
  {Poulin}}\ and\ \bibinfo {author} {\bibfnamefont {D.~A.}\ \bibnamefont
  {Weitz}},\ }\Doi {10.1103/PhysRevE.57.626} {\bibfield  {journal} {\bibinfo
  {journal} {Phys. Rev. E},\ }\textbf {\bibinfo {volume} {57}},\ \bibinfo
  {pages} {626} (\bibinfo {year} {1998})}\BibitemShut {NoStop}%
\bibitem [{\citenamefont {Araki}\ and\ \citenamefont
  {Tanaka}(2006)}]{PhysRevLett.97.127801}%
  \BibitemOpen
  \bibfield  {author} {\bibinfo {author} {\bibfnamefont {T.}~\bibnamefont
  {Araki}}\ and\ \bibinfo {author} {\bibfnamefont {H.}~\bibnamefont {Tanaka}},\
  }\Doi {10.1103/PhysRevLett.97.127801} {\bibfield  {journal} {\bibinfo
  {journal} {Phys. Rev. Lett.},\ }\textbf {\bibinfo {volume} {97}},\ \bibinfo
  {pages} {127801} (\bibinfo {year} {2006})}\BibitemShut {NoStop}%
\bibitem [{\citenamefont {Ravnik}\ \emph {et~al.}(2007)\citenamefont {Ravnik},
  \citenamefont {\ifmmode~\check{S}\else \v{S}\fi{}karabot}, \citenamefont
  {\ifmmode~\check{Z}\else \v{Z}\fi{}umer}, \citenamefont {Tkalec},
  \citenamefont {Poberaj}, \citenamefont {Babi\ifmmode~\check{c}\else
  \v{c}\fi{}}, \citenamefont {Osterman},\ and\ \citenamefont {Mu\ifmmode
  \check{s}\else \v{s}\fi{}evi\ifmmode~\check{c}\else
  \v{c}\fi{}}}]{PhysRevLett.99.247801}%
  \BibitemOpen
  \bibfield  {author} {\bibinfo {author} {\bibfnamefont {M.}~\bibnamefont
  {Ravnik}}, \bibinfo {author} {\bibfnamefont {M.}~\bibnamefont
  {\ifmmode~\check{S}\else \v{S}\fi{}karabot}}, \bibinfo {author}
  {\bibfnamefont {S.}~\bibnamefont {\ifmmode~\check{Z}\else \v{Z}\fi{}umer}},
  \bibinfo {author} {\bibfnamefont {U.}~\bibnamefont {Tkalec}}, \bibinfo
  {author} {\bibfnamefont {I.}~\bibnamefont {Poberaj}}, \bibinfo {author}
  {\bibfnamefont {D.}~\bibnamefont {Babi\ifmmode~\check{c}\else \v{c}\fi{}}},
  \bibinfo {author} {\bibfnamefont {N.}~\bibnamefont {Osterman}}, \ and\
  \bibinfo {author} {\bibfnamefont {I.}~\bibnamefont {Mu\ifmmode \check{s}\else
  \v{s}\fi{}evi\ifmmode~\check{c}\else \v{c}\fi{}}},\ }\Doi
  {10.1103/PhysRevLett.99.247801} {\bibfield  {journal} {\bibinfo  {journal}
  {Phys. Rev. Lett.},\ }\textbf {\bibinfo {volume} {99}},\ \bibinfo {pages}
  {247801} (\bibinfo {year} {2007})}\BibitemShut {NoStop}%
\bibitem [{\citenamefont {Tkalec}\ \emph {et~al.}(2011)\citenamefont {Tkalec},
  \citenamefont {Ravnik}, \citenamefont {{\v C}opar}, \citenamefont {{\v
  Z}umer},\ and\ \citenamefont {Mu{\v s}evi{\v c}}}]{Tkalec01072011}%
  \BibitemOpen
  \bibfield  {author} {\bibinfo {author} {\bibfnamefont {U.}~\bibnamefont
  {Tkalec}}, \bibinfo {author} {\bibfnamefont {M.}~\bibnamefont {Ravnik}},
  \bibinfo {author} {\bibfnamefont {S.}~\bibnamefont {{\v C}opar}}, \bibinfo
  {author} {\bibfnamefont {S.}~\bibnamefont {{\v Z}umer}}, \ and\ \bibinfo
  {author} {\bibfnamefont {I.}~\bibnamefont {Mu{\v s}evi{\v c}}},\ }\Doi
  {10.1126/science.1205705} {\bibfield  {journal} {\bibinfo  {journal}
  {Science},\ }\textbf {\bibinfo {volume} {333}},\ \bibinfo {pages} {62}
  (\bibinfo {year} {2011})}\BibitemShut {NoStop}%
\bibitem [{\citenamefont {Jampani}\ \emph {et~al.}(2011)\citenamefont
  {Jampani}, \citenamefont {\ifmmode~\check{S}\else \v{S}\fi{}karabot},
  \citenamefont {Ravnik}, \citenamefont {\ifmmode~\check{C}\else
  \v{C}\fi{}opar}, \citenamefont {\ifmmode~\check{Z}\else \v{Z}\fi{}umer},\
  and\ \citenamefont {Mu\ifmmode \check{s}\else
  \v{s}\fi{}evi\ifmmode~\check{c}\else \v{c}\fi{}}}]{PhysRevE.84.031703}%
  \BibitemOpen
  \bibfield  {author} {\bibinfo {author} {\bibfnamefont {V.~S.~R.}\
  \bibnamefont {Jampani}}, \bibinfo {author} {\bibfnamefont {M.}~\bibnamefont
  {\ifmmode~\check{S}\else \v{S}\fi{}karabot}}, \bibinfo {author}
  {\bibfnamefont {M.}~\bibnamefont {Ravnik}}, \bibinfo {author} {\bibfnamefont
  {S.}~\bibnamefont {\ifmmode~\check{C}\else \v{C}\fi{}opar}}, \bibinfo
  {author} {\bibfnamefont {S.}~\bibnamefont {\ifmmode~\check{Z}\else
  \v{Z}\fi{}umer}}, \ and\ \bibinfo {author} {\bibfnamefont {I.}~\bibnamefont
  {Mu\ifmmode \check{s}\else \v{s}\fi{}evi\ifmmode~\check{c}\else
  \v{c}\fi{}}},\ }\Doi {10.1103/PhysRevE.84.031703} {\bibfield  {journal}
  {\bibinfo  {journal} {Phys. Rev. E},\ }\textbf {\bibinfo {volume} {84}},\
  \bibinfo {pages} {031703} (\bibinfo {year} {2011})}\BibitemShut {NoStop}%
\bibitem [{\citenamefont {{Lopez-Leon}}\ \emph {et~al.}(2011)\citenamefont
  {{Lopez-Leon}}, \citenamefont {{Koning}}, \citenamefont {{Devaiah}},
  \citenamefont {{Vitelli}},\ and\ \citenamefont
  {{Fernandez-Nieves}}}]{2011NatPh...7..391L}%
  \BibitemOpen
  \bibfield  {author} {\bibinfo {author} {\bibfnamefont {T.}~\bibnamefont
  {{Lopez-Leon}}}, \bibinfo {author} {\bibfnamefont {V.}~\bibnamefont
  {{Koning}}}, \bibinfo {author} {\bibfnamefont {K.~B.~S.}\ \bibnamefont
  {{Devaiah}}}, \bibinfo {author} {\bibfnamefont {V.}~\bibnamefont
  {{Vitelli}}}, \ and\ \bibinfo {author} {\bibfnamefont {A.}~\bibnamefont
  {{Fernandez-Nieves}}},\ }\Doi {10.1038/nphys1920} {\bibfield  {journal}
  {\bibinfo  {journal} {Nature Physics},\ }\textbf {\bibinfo {volume} {7}},\
  \bibinfo {pages} {391} (\bibinfo {year} {2011})}\BibitemShut {NoStop}%
\bibitem [{\citenamefont {Koning}\ \emph {et~al.}(2013)\citenamefont {Koning},
  \citenamefont {Lopez-Leon}, \citenamefont {Fernandez-Nieves},\ and\
  \citenamefont {Vitelli}}]{C3SM27671F}%
  \BibitemOpen
  \bibfield  {author} {\bibinfo {author} {\bibfnamefont {V.}~\bibnamefont
  {Koning}}, \bibinfo {author} {\bibfnamefont {T.}~\bibnamefont {Lopez-Leon}},
  \bibinfo {author} {\bibfnamefont {A.}~\bibnamefont {Fernandez-Nieves}}, \
  and\ \bibinfo {author} {\bibfnamefont {V.}~\bibnamefont {Vitelli}},\ }\Doi
  {10.1039/C3SM27671F} {\bibfield  {journal} {\bibinfo  {journal} {Soft
  Matter},\ }\textbf {\bibinfo {volume} {9}},\ \bibinfo {pages} {4993}
  (\bibinfo {year} {2013})}\BibitemShut {NoStop}%
\bibitem [{\citenamefont {Senyuk}\ \emph {et~al.}(2013)\citenamefont {Senyuk},
  \citenamefont {Liu}, \citenamefont {He}, \citenamefont {Kamien},
  \citenamefont {Kusner}, \citenamefont {Lubensky},\ and\ \citenamefont
  {Smalyukh}}]{Senyuk:2013fk}%
  \BibitemOpen
  \bibfield  {author} {\bibinfo {author} {\bibfnamefont {B.}~\bibnamefont
  {Senyuk}}, \bibinfo {author} {\bibfnamefont {Q.}~\bibnamefont {Liu}},
  \bibinfo {author} {\bibfnamefont {S.}~\bibnamefont {He}}, \bibinfo {author}
  {\bibfnamefont {R.~D.}\ \bibnamefont {Kamien}}, \bibinfo {author}
  {\bibfnamefont {R.~B.}\ \bibnamefont {Kusner}}, \bibinfo {author}
  {\bibfnamefont {T.~C.}\ \bibnamefont {Lubensky}}, \ and\ \bibinfo {author}
  {\bibfnamefont {I.~I.}\ \bibnamefont {Smalyukh}},\ }\href
  {http://dx.doi.org/10.1038/nature11710} {\bibfield  {journal} {\bibinfo
  {journal} {Nature},\ }\textbf {\bibinfo {volume} {493}},\ \bibinfo {pages}
  {200} (\bibinfo {year} {2013})}\BibitemShut {NoStop}%
\bibitem [{\citenamefont {Liu}\ \emph {et~al.}(2013)\citenamefont {Liu},
  \citenamefont {Senyuk}, \citenamefont {Tasinkevych},\ and\ \citenamefont
  {Smalyukh}}]{Liu04062013}%
  \BibitemOpen
  \bibfield  {author} {\bibinfo {author} {\bibfnamefont {Q.}~\bibnamefont
  {Liu}}, \bibinfo {author} {\bibfnamefont {B.}~\bibnamefont {Senyuk}},
  \bibinfo {author} {\bibfnamefont {M.}~\bibnamefont {Tasinkevych}}, \ and\
  \bibinfo {author} {\bibfnamefont {I.~I.}\ \bibnamefont {Smalyukh}},\ }\Doi
  {10.1073/pnas.1301464110} {\bibfield  {journal} {\bibinfo  {journal}
  {Proceedings of the National Academy of Sciences},\ }\textbf {\bibinfo
  {volume} {110}},\ \bibinfo {pages} {9231} (\bibinfo {year}
  {2013})}\BibitemShut {NoStop}%
\bibitem [{\citenamefont {Cavallaro~Jr}\ \emph {et~al.}(2013)\citenamefont
  {Cavallaro~Jr}, \citenamefont {Gharbi}, \citenamefont {Beller}, \citenamefont
  {Copar}, \citenamefont {Shi}, \citenamefont {Kamien}, \citenamefont {Yang},
  \citenamefont {Baumgart},\ and\ \citenamefont {Stebe}}]{C3SM51167G}%
  \BibitemOpen
  \bibfield  {author} {\bibinfo {author} {\bibfnamefont {M.}~\bibnamefont
  {Cavallaro~Jr}}, \bibinfo {author} {\bibfnamefont {M.~A.}\ \bibnamefont
  {Gharbi}}, \bibinfo {author} {\bibfnamefont {D.~A.}\ \bibnamefont {Beller}},
  \bibinfo {author} {\bibfnamefont {S.}~\bibnamefont {Copar}}, \bibinfo
  {author} {\bibfnamefont {Z.}~\bibnamefont {Shi}}, \bibinfo {author}
  {\bibfnamefont {R.~D.}\ \bibnamefont {Kamien}}, \bibinfo {author}
  {\bibfnamefont {S.}~\bibnamefont {Yang}}, \bibinfo {author} {\bibfnamefont
  {T.}~\bibnamefont {Baumgart}}, \ and\ \bibinfo {author} {\bibfnamefont
  {K.~J.}\ \bibnamefont {Stebe}},\ }\Doi {10.1039/C3SM51167G} {\bibfield
  {journal} {\bibinfo  {journal} {Soft Matter},\ }\textbf {\bibinfo {volume}
  {9}},\ \bibinfo {pages} {9099} (\bibinfo {year} {2013})}\BibitemShut
  {NoStop}%
\bibitem [{\citenamefont {Stelzer}\ and\ \citenamefont
  {Bernhard}(2000)}]{cond-mat/0012394}%
  \BibitemOpen
  \bibfield  {author} {\bibinfo {author} {\bibfnamefont {J.}~\bibnamefont
  {Stelzer}}\ and\ \bibinfo {author} {\bibfnamefont {R.}~\bibnamefont
  {Bernhard}},\ }\href@noop {} { (\bibinfo {year} {2000})},\ \Eprint
  {http://arxiv.org/abs/cond-mat/0012394} {arXiv:cond-mat/0012394} \BibitemShut
  {NoStop}%
\bibitem [{\citenamefont {Bowick}\ \emph {et~al.}(2004)\citenamefont {Bowick},
  \citenamefont {Nelson},\ and\ \citenamefont
  {Travesset}}]{PhysRevE.69.041102}%
  \BibitemOpen
  \bibfield  {author} {\bibinfo {author} {\bibfnamefont {M.}~\bibnamefont
  {Bowick}}, \bibinfo {author} {\bibfnamefont {D.~R.}\ \bibnamefont {Nelson}},
  \ and\ \bibinfo {author} {\bibfnamefont {A.}~\bibnamefont {Travesset}},\
  }\Doi {10.1103/PhysRevE.69.041102} {\bibfield  {journal} {\bibinfo  {journal}
  {Phys. Rev. E},\ }\textbf {\bibinfo {volume} {69}},\ \bibinfo {pages}
  {041102} (\bibinfo {year} {2004})}\BibitemShut {NoStop}%
\bibitem [{\citenamefont {Kuli{\'c}}\ \emph {et~al.}(2004)\citenamefont
  {Kuli{\'c}}, \citenamefont {Andrienko},\ and\ \citenamefont
  {Deserno}}]{0295-5075-67-3-418}%
  \BibitemOpen
  \bibfield  {author} {\bibinfo {author} {\bibfnamefont {I.~M.}\ \bibnamefont
  {Kuli{\'c}}}, \bibinfo {author} {\bibfnamefont {D.}~\bibnamefont
  {Andrienko}}, \ and\ \bibinfo {author} {\bibfnamefont {M.}~\bibnamefont
  {Deserno}},\ }\href {http://stacks.iop.org/0295-5075/67/i=3/a=418} {\bibfield
   {journal} {\bibinfo  {journal} {EPL (Europhysics Letters)},\ }\textbf
  {\bibinfo {volume} {67}},\ \bibinfo {pages} {418} (\bibinfo {year}
  {2004})}\BibitemShut {NoStop}%
\bibitem [{\citenamefont {Giomi}\ and\ \citenamefont
  {Bowick}(2008)}]{PhysRevE.78.010601}%
  \BibitemOpen
  \bibfield  {author} {\bibinfo {author} {\bibfnamefont {L.}~\bibnamefont
  {Giomi}}\ and\ \bibinfo {author} {\bibfnamefont {M.~J.}\ \bibnamefont
  {Bowick}},\ }\Doi {10.1103/PhysRevE.78.010601} {\bibfield  {journal}
  {\bibinfo  {journal} {Phys. Rev. E},\ }\textbf {\bibinfo {volume} {78}},\
  \bibinfo {pages} {010601} (\bibinfo {year} {2008})}\BibitemShut {NoStop}%
\bibitem [{\citenamefont {{Giomi}}\ and\ \citenamefont
  {{Bowick}}(2008)}]{2008arXiv0807.4538G}%
  \BibitemOpen
  \bibfield  {author} {\bibinfo {author} {\bibfnamefont {L.}~\bibnamefont
  {{Giomi}}}\ and\ \bibinfo {author} {\bibfnamefont {M.~J.}\ \bibnamefont
  {{Bowick}}},\ }\href@noop {} {\bibfield  {journal} {\bibinfo  {journal} {Eur.
  Phys. J. E},\ }\textbf {\bibinfo {volume} {27}},\ \bibinfo {pages} {275}
  (\bibinfo {year} {2008})},\ \Eprint {http://arxiv.org/abs/0807.4538}
  {arXiv:0807.4538 [cond-mat.soft]} \BibitemShut {NoStop}%
\bibitem [{\citenamefont {{Bowick}}\ and\ \citenamefont
  {{Giomi}}(2009)}]{2009AdPhy..58..449B}%
  \BibitemOpen
  \bibfield  {author} {\bibinfo {author} {\bibfnamefont {M.}~\bibnamefont
  {{Bowick}}}\ and\ \bibinfo {author} {\bibfnamefont {L.}~\bibnamefont
  {{Giomi}}},\ }\Doi {10.1080/00018730903043166} {\bibfield  {journal}
  {\bibinfo  {journal} {Advances in Physics},\ }\textbf {\bibinfo {volume}
  {58}},\ \bibinfo {pages} {449} (\bibinfo {year} {2009})},\ \Eprint
  {http://arxiv.org/abs/0812.3064} {arXiv:0812.3064 [cond-mat.soft]}
  \BibitemShut {NoStop}%
\bibitem [{\citenamefont {Yao}\ and\ \citenamefont {de~la
  Cruz}(2013)}]{PhysRevE.87.012603}%
  \BibitemOpen
  \bibfield  {author} {\bibinfo {author} {\bibfnamefont {Z.}~\bibnamefont
  {Yao}}\ and\ \bibinfo {author} {\bibfnamefont {M.~O.}\ \bibnamefont {de~la
  Cruz}},\ }\Doi {10.1103/PhysRevE.87.012603} {\bibfield  {journal} {\bibinfo
  {journal} {Phys. Rev. E},\ }\textbf {\bibinfo {volume} {87}},\ \bibinfo
  {pages} {012603} (\bibinfo {year} {2013})}\BibitemShut {NoStop}%
\bibitem [{\citenamefont {Pairam}\ \emph {et~al.}(2013)\citenamefont {Pairam},
  \citenamefont {Vallamkondu}, \citenamefont {Koning}, \citenamefont {van
  Zuiden}, \citenamefont {Ellis}, \citenamefont {Bates}, \citenamefont
  {Vitelli},\ and\ \citenamefont {Fernandez-Nieves}}]{Pairam04062013}%
  \BibitemOpen
  \bibfield  {author} {\bibinfo {author} {\bibfnamefont {E.}~\bibnamefont
  {Pairam}}, \bibinfo {author} {\bibfnamefont {J.}~\bibnamefont {Vallamkondu}},
  \bibinfo {author} {\bibfnamefont {V.}~\bibnamefont {Koning}}, \bibinfo
  {author} {\bibfnamefont {B.~C.}\ \bibnamefont {van Zuiden}}, \bibinfo
  {author} {\bibfnamefont {P.~W.}\ \bibnamefont {Ellis}}, \bibinfo {author}
  {\bibfnamefont {M.~A.}\ \bibnamefont {Bates}}, \bibinfo {author}
  {\bibfnamefont {V.}~\bibnamefont {Vitelli}}, \ and\ \bibinfo {author}
  {\bibfnamefont {A.}~\bibnamefont {Fernandez-Nieves}},\ }\Doi
  {10.1073/pnas.1221380110} {\bibfield  {journal} {\bibinfo  {journal}
  {Proceedings of the National Academy of Sciences},\ }\textbf {\bibinfo
  {volume} {110}},\ \bibinfo {pages} {9295} (\bibinfo {year}
  {2013})}\BibitemShut {NoStop}%
\bibitem [{\citenamefont {Smalyukh}\ \emph {et~al.}(2010)\citenamefont
  {Smalyukh}, \citenamefont {Lansac}, \citenamefont {Clark},\ and\
  \citenamefont {Trivedi}}]{Smalyukh:2010fk}%
  \BibitemOpen
  \bibfield  {author} {\bibinfo {author} {\bibfnamefont {I.~I.}\ \bibnamefont
  {Smalyukh}}, \bibinfo {author} {\bibfnamefont {Y.}~\bibnamefont {Lansac}},
  \bibinfo {author} {\bibfnamefont {N.~A.}\ \bibnamefont {Clark}}, \ and\
  \bibinfo {author} {\bibfnamefont {R.~P.}\ \bibnamefont {Trivedi}},\ }\href
  {http://dx.doi.org/10.1038/nmat2592} {\bibfield  {journal} {\bibinfo
  {journal} {Nat Mater},\ }\textbf {\bibinfo {volume} {9}},\ \bibinfo {pages}
  {139} (\bibinfo {year} {2010})}\BibitemShut {NoStop}%
\bibitem [{Note1()}]{Note1}%
  \BibitemOpen
  \bibinfo {note} {Technically, it is spontaneous {\protect \sl achiral}
  symmetry breaking since the symmetry is the {\protect \sl lack} of chirality.
  However, we will conform to the standard convention.}\BibitemShut {Stop}%
\bibitem [{\citenamefont {Bloomfield}(1997)}]{BIP:BIP6}%
  \BibitemOpen
  \bibfield  {author} {\bibinfo {author} {\bibfnamefont {V.~A.}\ \bibnamefont
  {Bloomfield}},\ }\Doi
  {10.1002/(SICI)1097-0282(1997)44:3<269::AID-BIP6>3.0.CO;2-T} {\bibfield
  {journal} {\bibinfo  {journal} {Biopolymers},\ }\textbf {\bibinfo {volume}
  {44}},\ \bibinfo {pages} {269} (\bibinfo {year} {1997})},\ ISSN \bibinfo
  {issn} {1097-0282}\BibitemShut {NoStop}%
\bibitem [{\citenamefont {{Sven{\v s}ek}}\ and\ \citenamefont
  {{Podgornik}}(2012)}]{2012EL....10066005S}%
  \BibitemOpen
  \bibfield  {author} {\bibinfo {author} {\bibfnamefont {D.}~\bibnamefont
  {{Sven{\v s}ek}}}\ and\ \bibinfo {author} {\bibfnamefont {R.}~\bibnamefont
  {{Podgornik}}},\ }\Doi {10.1209/0295-5075/100/66005} {\bibfield  {journal}
  {\bibinfo  {journal} {EPL (Europhysics Letters)},\ }\textbf {\bibinfo
  {volume} {100}},\ \bibinfo {pages} {66005} (\bibinfo {year} {2012})},\
  \Eprint {http://arxiv.org/abs/1210.3228} {arXiv:1210.3228 [cond-mat.soft]}
  \BibitemShut {NoStop}%
\bibitem [{\citenamefont {Lifshitz}\ \emph {et~al.}(1978)\citenamefont
  {Lifshitz}, \citenamefont {Grosberg},\ and\ \citenamefont
  {Khokhlov}}]{RevModPhys.50.683}%
  \BibitemOpen
  \bibfield  {author} {\bibinfo {author} {\bibfnamefont {I.~M.}\ \bibnamefont
  {Lifshitz}}, \bibinfo {author} {\bibfnamefont {A.~Y.}\ \bibnamefont
  {Grosberg}}, \ and\ \bibinfo {author} {\bibfnamefont {A.~R.}\ \bibnamefont
  {Khokhlov}},\ }\Doi {10.1103/RevModPhys.50.683} {\bibfield  {journal}
  {\bibinfo  {journal} {Rev. Mod. Phys.},\ }\textbf {\bibinfo {volume} {50}},\
  \bibinfo {pages} {683} (\bibinfo {year} {1978})}\BibitemShut {NoStop}%
\bibitem [{\citenamefont {{Stevens}}(2001)}]{2001BpJ....80..130S}%
  \BibitemOpen
  \bibfield  {author} {\bibinfo {author} {\bibfnamefont {M.}~\bibnamefont
  {{Stevens}}},\ }\Doi {10.1016/S0006-3495(01)76000-6} {\bibfield  {journal}
  {\bibinfo  {journal} {Biophysical Journal},\ }\textbf {\bibinfo {volume}
  {80}},\ \bibinfo {pages} {130} (\bibinfo {year} {2001})}\BibitemShut
  {NoStop}%
\bibitem [{\citenamefont {{Stukan}}\ \emph {et~al.}(2003)\citenamefont
  {{Stukan}}, \citenamefont {{Ivanov}}, \citenamefont {{Grosberg}},
  \citenamefont {{Paul}},\ and\ \citenamefont
  {{Binder}}}]{2003JChPh.118.3392S}%
  \BibitemOpen
  \bibfield  {author} {\bibinfo {author} {\bibfnamefont {M.~R.}\ \bibnamefont
  {{Stukan}}}, \bibinfo {author} {\bibfnamefont {V.~A.}\ \bibnamefont
  {{Ivanov}}}, \bibinfo {author} {\bibfnamefont {A.~Y.}\ \bibnamefont
  {{Grosberg}}}, \bibinfo {author} {\bibfnamefont {W.}~\bibnamefont {{Paul}}},
  \ and\ \bibinfo {author} {\bibfnamefont {K.}~\bibnamefont {{Binder}}},\ }\Doi
  {10.1063/1.1536620} {\bibfield  {journal} {\bibinfo  {journal} {\jcp},\
  }\textbf {\bibinfo {volume} {118}},\ \bibinfo {pages} {3392} (\bibinfo {year}
  {2003})}\BibitemShut {NoStop}%
\bibitem [{\citenamefont {{Conwell}}\ \emph {et~al.}(2003)\citenamefont
  {{Conwell}}, \citenamefont {{Vilfan}},\ and\ \citenamefont
  {{Hud}}}]{2003PNAS..100.9296C}%
  \BibitemOpen
  \bibfield  {author} {\bibinfo {author} {\bibfnamefont {C.~C.}\ \bibnamefont
  {{Conwell}}}, \bibinfo {author} {\bibfnamefont {I.~D.}\ \bibnamefont
  {{Vilfan}}}, \ and\ \bibinfo {author} {\bibfnamefont {N.~V.}\ \bibnamefont
  {{Hud}}},\ }\Doi {10.1073/pnas.1533135100} {\bibfield  {journal} {\bibinfo
  {journal} {Proceedings of the National Academy of Science},\ }\textbf
  {\bibinfo {volume} {100}},\ \bibinfo {pages} {9296} (\bibinfo {year}
  {2003})}\BibitemShut {NoStop}%
\bibitem [{\citenamefont {de~Gennes}\ and\ \citenamefont
  {Prost}(1993)}]{deGennes}%
  \BibitemOpen
  \bibfield  {author} {\bibinfo {author} {\bibfnamefont {P.~G.}\ \bibnamefont
  {de~Gennes}}\ and\ \bibinfo {author} {\bibfnamefont {J.}~\bibnamefont
  {Prost}},\ }\href@noop {} {\emph {\bibinfo {title} {The Physics of Liquid
  Crystals}}}\ (\bibinfo  {publisher} {Oxford University Press, New York},\
  \bibinfo {year} {1993})\BibitemShut {NoStop}%
\bibitem [{\citenamefont {Kleman}\ and\ \citenamefont
  {Lavrentovich}(2003)}]{SoftMatter}%
  \BibitemOpen
  \bibfield  {author} {\bibinfo {author} {\bibfnamefont {M.}~\bibnamefont
  {Kleman}}\ and\ \bibinfo {author} {\bibfnamefont {O.~D.}\ \bibnamefont
  {Lavrentovich}},\ }\href@noop {} {\emph {\bibinfo {title} {Soft Matter
  Physics: An Introduction}}}\ (\bibinfo  {publisher} {Springer-Verlag New
  York, Inc.},\ \bibinfo {year} {2003})\BibitemShut {NoStop}%
\bibitem [{\citenamefont {Kamien}(2002)}]{RevModPhys.74.953}%
  \BibitemOpen
  \bibfield  {author} {\bibinfo {author} {\bibfnamefont {R.~D.}\ \bibnamefont
  {Kamien}},\ }\Doi {10.1103/RevModPhys.74.953} {\bibfield  {journal} {\bibinfo
   {journal} {Rev. Mod. Phys.},\ }\textbf {\bibinfo {volume} {74}},\ \bibinfo
  {pages} {953} (\bibinfo {year} {2002})}\BibitemShut {NoStop}%
\bibitem [{Note2()}]{Note2}%
  \BibitemOpen
  \bibinfo {note} {Explicitly: $\protect \{K_i\protect \}=\protect
  \{K_1,K_2,K_3,K_{24}\protect \}$}\BibitemShut {NoStop}%
\bibitem [{Note3()}]{Note3}%
  \BibitemOpen
  \bibinfo {note} {The fourth order term in the bend energy for general $\xi $,
  that reduces to $\protect \frac { \pi ^2}{2} K_3 R_2 \xi \omega ^4$ in eq.
  \ref {eq:largexi}, is not given in eq. \protect \textup {\hbox {\mathsurround
  \z@ \protect \normalfont (\ignorespaces \ref {eq:bend}\unskip \@@italiccorr
  )}}, because it is too lengthy.}\BibitemShut {Stop}%
\bibitem [{\citenamefont {Meyer}(1969)}]{PhysRevLett.22.918}%
  \BibitemOpen
  \bibfield  {author} {\bibinfo {author} {\bibfnamefont {R.~B.}\ \bibnamefont
  {Meyer}},\ }\Doi {10.1103/PhysRevLett.22.918} {\bibfield  {journal} {\bibinfo
   {journal} {Phys. Rev. Lett.},\ }\textbf {\bibinfo {volume} {22}},\ \bibinfo
  {pages} {918} (\bibinfo {year} {1969})}\BibitemShut {NoStop}%
\bibitem [{\citenamefont {Fornberg}(1988)}]{fornberg1988generation}%
  \BibitemOpen
  \bibfield  {author} {\bibinfo {author} {\bibfnamefont {B.}~\bibnamefont
  {Fornberg}},\ }\href@noop {} {\bibfield  {journal} {\bibinfo  {journal}
  {Mathematics of Computation},\ }\textbf {\bibinfo {volume} {51}},\ \bibinfo
  {pages} {699} (\bibinfo {year} {1988})}\BibitemShut {NoStop}%
\bibitem [{\citenamefont {Allender}\ \emph {et~al.}(1991)\citenamefont
  {Allender}, \citenamefont {Crawford},\ and\ \citenamefont
  {Doane}}]{PhysRevLett.67.1442}%
  \BibitemOpen
  \bibfield  {author} {\bibinfo {author} {\bibfnamefont {D.~W.}\ \bibnamefont
  {Allender}}, \bibinfo {author} {\bibfnamefont {G.~P.}\ \bibnamefont
  {Crawford}}, \ and\ \bibinfo {author} {\bibfnamefont {J.~W.}\ \bibnamefont
  {Doane}},\ }\Doi {10.1103/PhysRevLett.67.1442} {\bibfield  {journal}
  {\bibinfo  {journal} {Phys. Rev. Lett.},\ }\textbf {\bibinfo {volume} {67}},\
  \bibinfo {pages} {1442} (\bibinfo {year} {1991})}\BibitemShut {NoStop}%
\bibitem [{\citenamefont {Lavrentovich}\ and\ \citenamefont
  {Pergamenshchik}(1994)}]{PhysRevLett.73.979}%
  \BibitemOpen
  \bibfield  {author} {\bibinfo {author} {\bibfnamefont {O.~D.}\ \bibnamefont
  {Lavrentovich}}\ and\ \bibinfo {author} {\bibfnamefont {V.~M.}\ \bibnamefont
  {Pergamenshchik}},\ }\Doi {10.1103/PhysRevLett.73.979} {\bibfield  {journal}
  {\bibinfo  {journal} {Phys. Rev. Lett.},\ }\textbf {\bibinfo {volume} {73}},\
  \bibinfo {pages} {979} (\bibinfo {year} {1994})}\BibitemShut {NoStop}%
\bibitem [{\citenamefont {Polak}\ \emph {et~al.}(1994)\citenamefont {Polak},
  \citenamefont {Crawford}, \citenamefont {Kostival}, \citenamefont {Doane},\
  and\ \citenamefont {\ifmmode~\check{Z}\else
  \v{Z}\fi{}umer}}]{PhysRevE.49.R978}%
  \BibitemOpen
  \bibfield  {author} {\bibinfo {author} {\bibfnamefont {R.~D.}\ \bibnamefont
  {Polak}}, \bibinfo {author} {\bibfnamefont {G.~P.}\ \bibnamefont {Crawford}},
  \bibinfo {author} {\bibfnamefont {B.~C.}\ \bibnamefont {Kostival}}, \bibinfo
  {author} {\bibfnamefont {J.~W.}\ \bibnamefont {Doane}}, \ and\ \bibinfo
  {author} {\bibfnamefont {S.}~\bibnamefont {\ifmmode~\check{Z}\else
  \v{Z}\fi{}umer}},\ }\Doi {10.1103/PhysRevE.49.R978} {\bibfield  {journal}
  {\bibinfo  {journal} {Phys. Rev. E},\ }\textbf {\bibinfo {volume} {49}},\
  \bibinfo {pages} {R978} (\bibinfo {year} {1994})}\BibitemShut {NoStop}%
\bibitem [{\citenamefont {Sparavigna}\ \emph {et~al.}(1994)\citenamefont
  {Sparavigna}, \citenamefont {Lavrentovich},\ and\ \citenamefont
  {Strigazzi}}]{PhysRevE.49.1344}%
  \BibitemOpen
  \bibfield  {author} {\bibinfo {author} {\bibfnamefont {A.}~\bibnamefont
  {Sparavigna}}, \bibinfo {author} {\bibfnamefont {O.~D.}\ \bibnamefont
  {Lavrentovich}}, \ and\ \bibinfo {author} {\bibfnamefont {A.}~\bibnamefont
  {Strigazzi}},\ }\Doi {10.1103/PhysRevE.49.1344} {\bibfield  {journal}
  {\bibinfo  {journal} {Phys. Rev. E},\ }\textbf {\bibinfo {volume} {49}},\
  \bibinfo {pages} {1344} (\bibinfo {year} {1994})}\BibitemShut {NoStop}%
\bibitem [{\citenamefont {Lavrentovich}\ and\ \citenamefont
  {Pergamenshchik}(1995)}]{doi:10.1142/S0217979295000926}%
  \BibitemOpen
  \bibfield  {author} {\bibinfo {author} {\bibfnamefont {O.}~\bibnamefont
  {Lavrentovich}}\ and\ \bibinfo {author} {\bibfnamefont {V.}~\bibnamefont
  {Pergamenshchik}},\ }\Doi {10.1142/S0217979295000926} {\bibfield  {journal}
  {\bibinfo  {journal} {International Journal of Modern Physics B},\ }\textbf
  {\bibinfo {volume} {09}},\ \bibinfo {pages} {2389} (\bibinfo {year}
  {1995})}\BibitemShut {NoStop}%
\bibitem [{\citenamefont {Prinsen}\ and\ \citenamefont {van~der
  Schoot}(2004)}]{0953-8984-16-49-003}%
  \BibitemOpen
  \bibfield  {author} {\bibinfo {author} {\bibfnamefont {P.}~\bibnamefont
  {Prinsen}}\ and\ \bibinfo {author} {\bibfnamefont {P.}~\bibnamefont {van~der
  Schoot}},\ }\href {http://stacks.iop.org/0953-8984/16/i=49/a=003} {\bibfield
  {journal} {\bibinfo  {journal} {Journal of Physics: Condensed Matter},\
  }\textbf {\bibinfo {volume} {16}},\ \bibinfo {pages} {8835} (\bibinfo {year}
  {2004})}\BibitemShut {NoStop}%
\bibitem [{\citenamefont {Tortora}\ and\ \citenamefont
  {Lavrentovich}(2011)}]{Tortora29032011}%
  \BibitemOpen
  \bibfield  {author} {\bibinfo {author} {\bibfnamefont {L.}~\bibnamefont
  {Tortora}}\ and\ \bibinfo {author} {\bibfnamefont {O.~D.}\ \bibnamefont
  {Lavrentovich}},\ }\Doi {10.1073/pnas.1100087108} {\bibfield  {journal}
  {\bibinfo  {journal} {Proceedings of the National Academy of Sciences},\
  }\textbf {\bibinfo {volume} {108}},\ \bibinfo {pages} {5163} (\bibinfo {year}
  {2011})}\BibitemShut {NoStop}%
\bibitem [{\citenamefont {Lavrentovich}\ and\ \citenamefont
  {Sergan}(1990)}]{Lavrentovich:1990fk}%
  \BibitemOpen
  \bibfield  {author} {\bibinfo {author} {\bibfnamefont {O.}~\bibnamefont
  {Lavrentovich}}\ and\ \bibinfo {author} {\bibfnamefont {V.}~\bibnamefont
  {Sergan}},\ }\Doi {10.1007/BF02450386} {\bibfield  {journal} {\bibinfo
  {journal} {Il Nuovo Cimento D},\ }\textbf {\bibinfo {volume} {12}},\ \bibinfo
  {pages} {1219} (\bibinfo {year} {1990})},\ ISSN \bibinfo {issn}
  {0392-6737}\BibitemShut {NoStop}%
\bibitem [{\citenamefont {Williams}(1986)}]{0305-4470-19-16-019}%
  \BibitemOpen
  \bibfield  {author} {\bibinfo {author} {\bibfnamefont {R.~D.}\ \bibnamefont
  {Williams}},\ }\href {http://stacks.iop.org/0305-4470/19/i=16/a=019}
  {\bibfield  {journal} {\bibinfo  {journal} {Journal of Physics A:
  Mathematical and General},\ }\textbf {\bibinfo {volume} {19}},\ \bibinfo
  {pages} {3211} (\bibinfo {year} {1986})}\BibitemShut {NoStop}%
\end{thebibliography}%
\end{document}